\documentclass[a4paper,fleqn,usenatbib]{mnras}

\usepackage{amsmath}	% Advanced maths commands

\usepackage[T1]{fontenc}
\usepackage{ae,aecompl}

\usepackage{graphicx}	% Including figure files

\usepackage{amssymb}	% Extra maths symbols

\usepackage{booktabs}
\usepackage{longtable}

\title[Dust chemical evolution]{A new galactic chemical evolution model with dust: results for dwarf irregular galaxies and DLA systems}
\author[L. Gioannini et al.]{L. Gioannini$^{1}$\thanks{E-mail:
gioannini@oats.inaf.it}, F. Matteucci$^{1,2,3}$, G. Vladilo$^2$, F. Calura$^4$
\\
$^{1}$Dipartimento di Fisica, Sezione di Astronomia, Universit\`a di Trieste, via G.B. Tiepolo 11, 34100, Trieste, Italy\\
$^{2}$INAF, Osservatorio Astronomico di Trieste, via G.B. Tiepolo 11, 34100, Trieste, Italy\\
$^{3}$INFN, Sezione di Trieste, Via Valerio 2, 34100, Trieste, Italy\\
$^{4}$INAF, Osservatorio Astronomico di Bologna, via Ranzani 1, I-40127 Bologna, Italy 
}

\begin{document}

% \date{Accepted 2015 November 03. Received 2015 November 03; in original form 2015 March 26}

\pagerange{\pageref{firstpage}--\pageref{lastpage}} \pubyear{2016}

\maketitle

\label{firstpage}

\begin{abstract}
We present a galactic chemical evolution model which adopts updated prescriptions 
for all the main processes governing the dust cycle. We follow in detail the evolution of the abundances of several chemical species (C, O, S, Si, Fe and Zn) in the gas and dust of a typical dwarf irregular galaxy. The dwarf irregular galaxy is assumed to evolve with a low but continuous level of star formation and experience galactic winds triggered by supernova explosions. 
We predict the evolution of the gas to dust ratio in such a galaxy and discuss critically the main processes involving dust, such as dust production by AGB stars and Type II SNe, destruction and accretion (gas condensation in clouds). We then apply our model to Damped Lyman-$\alpha$ systems which are believed to be dwarf irregulars, as witnessed by their abundance patterns.
Our main conclusions are: i) we can reproduce the observed gas to dust ratio in dwarf galaxies. ii) We find that the process of dust accretion plays a fundamental role in the evolution of dust and in certain cases it becomes the dominant process in the dust cycle. On the other hand, dust destruction seems to be a negligible process in irregulars. iii) Concerning Damped Lyman-$\alpha$ systems, we show that the observed gas-phase abundances of silicon, normalized  to volatile elements (zinc and sulfur), are in agreement with our model. iv) The abundances of iron and silicon in DLA systems suggest that the two elements undergo a different history of dust formation and evolution. Our
work casts light on the nature of iron-rich dust: the observed depletion pattern of iron is well reproduced only when an additional source of iron dust is considered. 
Here we explore the possibility of a contribution from Type Ia SNe as well as an efficient accretion of iron nano-particles.
\end{abstract}


\section{Introduction}\label{introduction} 
The origin and the evolution of dust is one of the most important problems in Astrophysics. Cosmic dust plays a central role in the physics of the interstellar medium (ISM): it governs the scattering, absorption, re-emission of stellar light (Des\'ert et al. 1990; Witt $\&$ Gordon 2000) and it affects the spectral energy distribution (SED) of background sources (Silva et al. 1999; Granato et al. 2000). 
Dust properties have been determined from many kind of observations such as infrared continuum emission, depletion patterns in the ISM (Jenkins et al. 2009), isotopic anomalies in meteorites (Gail et al. 2009), extinction (Aguirre et al. 1999) etc.
Refractory elements (e.g., Si, Mg, Fe, Ni) are the ones which are subject to elemental depletion since a fraction of their abundances in the ISM is incorporated into dust grains. The circumstellar environments of evolved stars represents the sites where cosmic dust comes from, producing materials of silicate and carbonaceous type, i.e. the most important populations of dust species in the Universe (Draine $\&$ Li 2007). Stellar winds eject these dust particles in the ISM, and then, dust experiences lots of processes, which can decrease or increase its abundance and affect its size (Ferrarotti $\&$ Gail 2006, Zhukovska $\&$ Gail 2008). Thermal sputtering, evaporation in grain-grain collision, thermal sublimation or desorption are some examples of destruction processes, but the most important mechanism for cycling dust back to the gas phase resides in supernova shocks (Jones et al. 1994, McKee 1989). On the other hand, grain growth by dust coagulation and metals accretion onto preexisting grains increases 
either the dust mass or size and preferably occur in molecular clouds (Liffman $\&$ Clayton 1989, Hirashita 2000, Asano et al. 2013). These clouds are the sites where stars form and where new production of dust occur. All these processes together give rise to the so called ``dust cycle''.

Dwek (1998; hereafter D98) developed a chemical evolution model of the Milky Way, taking into account all the processes participating in the dust cycle. Since D98, significant progress  has been made concerning dust properties, both in theory and in observations. Calura et al. (2008; hereafter C08) modeled the evolution of dust in galaxies of different morphological types. New theoretical prescriptions about dust processing have appeared in more recent papers (Inoue 2011; Piovan et al. 2011; Asano et al. 2013; Hirashita 2013; Mattsson et al. 2015). 
High quality observations carried out using satellites and ground based telescopes have shed light on the nature and composition of the dust in local and high-redshift galaxies (Carilli et al. 2001; Draine 2003; Micha{\l}owski et al. 2010; Wang et al. 2010; Gall et al. 2011).
In particular, Damped Lyman Alpha (DLA) systems (Wolfe et al. 1986, 2005) offer a great opportunity for studying the composition of the ISM and  constraining dust properties at different cosmic times (Pei et al. 1991; Pettini et al. 1994; Vladilo $\&$ P\'eroux 2005; Vladilo et al. 2011).

In this work, we present a galactic chemical evolution model that incorporates updated prescriptions for dust production, accretion and destruction. We compare our results with other models widely employed in literature, and we constrain the origin and properties of cosmic dust by comparing these models with data of dwarf irregular galaxies and DLA systems. One of the specific aims is to find a plausible interpretation, in terms of dust evolution, of the  rise of iron depletion with increasing metallicity in DLA systems that has been known for a long time but for which there are no clear explanations (Vladilo 2004).

In the first part of the paper we present the new chemical model with dust, which adopts updated prescriptions for all the main processes governing the dust cycle. In section~\ref{chemical_model} we present the chemical evolution model adopted while the explanation for the dust model will be given in section~\ref{dust_model}. In section~\ref{dust_model_comparison} we present our results on the amount, composition and evolution of dust in dwarf irregular galaxies.
In the second part we show in section~\ref{DLA_sec} the comparison between our dust model and observational data of DLA systems. Finally, in section~\ref{sec_conclusions} some conclusions are drawn.

\section{Chemical evolution model} \label{chemical_model}
To study the evolution of chemical abundances we use self-consistent chemical evolution models, in which the instantaneous recycling approximation (IRA) is relaxed and the stellar lifetimes are taken into account. 

The model assumes that dwarf galaxies form by the infall of primordial gas (\textit{infall mass} $M_{infall}$), which accumulates in a pre-existing dark matter halo. Dwarf galaxies are described in more detail in Bradamante et al. (1998) and Lanfranchi $\&$ Matteucci (2004).

The birthrate function represents the number of stars formed in the mass interval $m$ and $m+dm$ in the time range between $t$ and $t + dt$. It depends on two physical quantities, the star formation rate ($\textit{SFR}=\psi(t)$) and the stellar initial mass function (\textit{IMF}=$\phi(m)$):
\begin{equation}
 B(m,t)=\psi(t)\phi(m)
\end{equation}
The star formation rate determines the rate at which the stars form and it is usually expressed in solar masses per year. We adopt a simple Schmidt law for the SFR:
\begin{equation}
 \psi(t)=\nu G(t)^k,
\end{equation}
where $\nu$ is the \textit{star formation efficiency} [$G yr^{-1}$] and the parameter $k$ is set equal to 1. $G(t)$ is the mass fraction of the ISM relative to the total mass accumulated up to the present time, $G(t)=M_{ISM}(t)/M_{tot}(t_{G})$. The star formation efficiency, defined as the SFR per unit mass of gas, can assume very different values depending on the morphological type of the modeled galaxy. Its inverse represents the time scale at which the total amount of gas is converted into stars. 
In this work, the initial mass function is assumed to be constant in space and time and normalized to unity in the mass interval between 0.1 and 100 $M_{\odot}$. In this work, we will adopt the Salpeter (1995) IMF:
\begin{equation}
 \phi_{Salp} (m)\propto m^{-(1+1.35)}  
 \end{equation}
and the Scalo (1986) IMF, characterized by a two slope power law: 
 \begin{equation}
\phi_{Scalo}(m)=
\begin{cases} 0.19 \cdot m^{-(1+1.35)}, &  for \,\,\,\, m<2 M_{\odot}  \\ \\
0.24 \cdot m^{-(1+1.70)}, &  for \,\,\,\, m>2 M_{\odot}  
\end{cases}
\end{equation}

In the next section we will show and discuss the basic equations used in the chemical evolution code which take into account the evolution of stars, SNe feedback, galactic winds and the infall of primordial gas.

\subsection{Basic equations}
Let us define $G_{i}(t)=G(t)X_{i}(t)$ as the fractional mass of the element $i$ at the time $t$ in the ISM, where $X_i(t)$ represents the abundance of the element $i$ in the gas at the time $t$. The temporal evolution of $G_{i}(t)$ is described by the following expression: 
\begin{equation} \label{basic_model}
\dot {G}_{i}(t) = -\psi(t)X_i(t) + R_i(t) + \dot {G}_{i,inf}(t) - \dot {G}_{i,w}(t),
\end{equation}
\begin{enumerate}
 \item The first term represents the rate at which the fraction of the element $i$ is removed by the ISM due to the SFR.
 \item $R_i(t)$ is the returned mass fraction of the element $i$ injected into the ISM from stars thanks to stellar winds and SN explosions. This term takes into account nucleosynthesis prescriptions concerning stellar yields and supernova progenitor models. $R_i(t)$ can be described as in Matteucci $\&$ Greggio (1986):

 \begin{multline} \label{R(t)}
R_{i}(t) = + \int_{M_{L}}^{M_{B_m}}\psi(t-\tau_m) Q_{mi}(t-\tau_m)\phi(m)dm +\\   
 A\int\limits_{M_{B_m}}^{M_{B_M}} \phi(m) \cdot
  \left[\,\,\int\limits_{\mu_{min}}^{0.5}f(\mu)\psi(t-\tau_{m2}) Q_{mi}(t-\tau_{m2})d\mu\right]dm \\  
 + (1-A)\int_{M_{B_m}}^{M_{B_M}}\psi(t-\tau_{m})  Q_{mi}(t-\tau_m)\phi(m)dm\\ 
 + \int_{M_{B_M}}^{M_U}\psi(t-\tau_m) Q_{mi}(t-\tau_m) \phi(m)dm \\ 
\end{multline}

The first term on the right side takes into account the enrichment of the element $i$ restored in the ISM by individual stars with a mass range between $M_{L}$ and $M_{B_m}$, which are respectively the minimum mass of a star contributing to the chemical enrichment of the ISM ($M_L =0.8 M_{\odot}$) and the minimum mass of a binary system that can give rise to Type Ia SNe ($M_{B_m}=3 M_{\odot}$). $Q_{mi}(t-\tau_m)$, where $\tau_{m}$ is the lifetime of a star of mass $m$, contains all the information about stellar nucleosynthesis for elements produced or destroyed by nuclear reactions inside each single star and injected into the ISM (Talbot $\&$ Arnett 1971). 
The second term gives the enrichment due to binary systems originating Type Ia SNe. For this type of SNe the single degenerate scenario is assumed, where a single C-O white dwarf explodes by the C-deflagration mechanism after having exceeded the Chandrasekhar mass ($1.44 M_{\odot}$). $A$ is a parameter representing the unknown fraction of binary stars giving rise to Type Ia SNe and it is fixed by reproducing the observed present time SN Ia rate. $\mu=M_2/M_B$ is defined as the ratio between the secondary component of the system over the total mass and $f(\mu)$ represents the distribution of this ratio. $M_{B_M}$ is the mass limit of the system and it is set to $16 M_{\odot}$: in fact, if one of the two stars in the system exceeds the mass of $8 M_{\odot}$, a Type II SN would result. Finally, $\tau_{m2}$ is the lifetime of the secondary star of the binary system, or rather the explosion timescale. 
The third term of Eq.(\ref{R(t)}) represents the enrichment due to stars in the mass range $M_{B_m} - M_{B_M}$, which are single, or if in binaries, do not produce an explosion of SN Ia. In this mass range, stars with $m>8M_{\odot}$ explode as Type II SNe. The fourth term of Eq.(\ref{R(t)}) concerns the stars with masses above $M_{B_M}$ and lower than $100 M_{\odot}$: all these stars explode as core-collapse SNe. 
\item The third term of Eq.(\ref{basic_model}) represents the rate of the infall of the element $i$. The infalling gas is not pre-enriched and consists in a pure primordial composition. The infall rate follows a decaying exponential law characterized by the \textit{infall time-scale} $\tau$: 
\begin{equation} 
\dot {G}_{i,inf}(t)= {\Gamma X_{i,inf} e^{-t/ \tau} \over M_{tot}(t_G)}
\end{equation}
where $\Gamma$ is the normalization constant constrained to reproduce the total mass at the present time. The time-scale of the infall $\tau$ is a free parameter which is fixed by reproducing the observed infall rate of the studied galaxy.
\item The last term of Eq.(\ref{basic_model}) concerns the outflow of the element $i$ due to galactic wind which occurs when the thermal energy of the gas heated by SN explosions exceeds its binding energy. The rate of the gas lost via galactic wind is proportional to the SFR and it is described as follows: 
\begin{equation} 
\dot{G}_{i,w}(t)= \omega_{i} \psi(t),
\end{equation}
where $\omega_i$ is a parameter which in our model is set equal for each element or, in other words, the wind is not differential.
\end{enumerate}

\subsubsection{Nucleosynthetic prescriptions}
Stars reprocess the ISM and contribute to the chemical enrichment of a galaxy. The stellar yields represent the amount of both newly formed and pre-existing elements injected into the ISM by the stars when they die. In our model we adopt different stellar yields for the contribution of low mass stars, Type Ia and Type II SNe:

\begin{enumerate}
 \item For low and intermediate mass stars (\textit{LIMS} with masses $0.8M_{\odot} < m < 8M_{\odot}$) we use the metallicity-dependent yields of van den Hoek $\&$ Groenewegen (1997).
 \item We assume that massive stars ($m>8M_{\odot}$) explode as core collapse SNe adopting the yields suggested by Francois et al. (2004), who performed an empirical modification to the ones of Woosley $\&$ Weaver (1995). In the case of sulfur we adopted the yields suggested in Vladilo et al. (2011).
 \item For Type Ia SNe we assume the yields of Iwamoto et al. (1999).

\end{enumerate}

\section{Dust chemical evolution model} \label{dust_model}
The evolution of dust is one of the most critical issues in Astrophysics. It is first produced by different types of stars, but during the galactic evolution, several physical phenomena critically modify the ISM and therefore the interstellar dust. In particular, astronomical observations and pre-solar grains analysis from meteorites indicate that physical processes responsible for the evolution of dust can be divided in two groups. The first takes into account all the mechanisms which change the total dust mass (destruction processes and grain growth by accretion), while the second group includes processes, like shattering and coagulation (Asano et al. 2013), which affect the grain size distribution. 
This work considers the evolution of the total amount of dust and its chemical composition during the cosmic time, so all the processes affecting the grain-size distribution are not taken into account.

We use the same approach first used by D98, later adopted in C08 and more recently by Grieco et al. (2014). The equation for the dust evolution is similar to Eq.(\ref{chemical_model}), but it includes all the physical processes which change the mass distribution of dust in the ISM, beside dust production by stars. Defining $G_{i,dust}=X_{i,dust}\cdot G(t)$ as the normalized mass of the element $i$ at the time $t$ in the dust phase, we can write:
\begin{equation} \label{chem-DUST}
\begin{split}
\dot {G}_{i,dust}(t)& = -\psi(t)X_{i,dust}(t) + R_{i,dust}(t)  + \left( \dfrac{{G}_{i,dust}(t)}{\tau_{accr}}\right) \\ 
& \quad - \left( \dfrac{{G}_{i,dust}(t)}{\tau_{destr}} \right)  \quad - \dot {G}_{i,dust}(t)_{w}, \\
\end{split}
\end{equation}
This equation takes into account all the processes which govern the so called dust cycle: the first term on the right side of the equation represents the rate at which the dust is removed from the ISM and utilized to form new stars (astration), whereas the second one gives the dust production rate of stars; the third and the fourth terms represent the processes which occur in the ISM and are dust accretion and destruction, respectively; the last term indicates the rate of dust expelled by galactic winds assuming dust and ISM to be coupled. In the next paragraphs we will discuss these terms in more details (see also C08 for a more detailed description).

\subsection{Dust formation} \label{sformation}
AGB stars and SNe represent the first environments where dust form: depending on the physical structure of the progenitor, various dust species can originate. 

The second term in the right side of Eq.(\ref{chem-DUST}) deals with the dust production by stars and can be described by the following expression:

\begin{multline}\label{R(t)dust}
 R_{i,dust}(t) =  \\
 + \int_{M_{L}}^{M_{B_m}}\psi(t-\tau_m) \delta^{AGB}_{i} Q_{mi}(t-\tau_m)\phi(m)dm   \\ 
 + A\int\limits_{M_	{B_m}}^{M_{B_M}} \phi(m)\cdot
 \left[\,\,\int\limits_{\mu_{min}}^{0.5}f(\mu)\psi(t-\tau_{m2})\delta^{Ia}_{i} Q_{mi}(t-\tau_{m2})d\mu\right]dm \\ 
 +(1-A)\int_{M_{B_m}}^{8 M_{\odot}}\psi(t-\tau_{m})  \delta^{AGB}_{i} Q_{mi}(t-\tau_m)\phi(m)dm \\
  +(1-A)\int_{8 M_{\odot}}^{M_{B_M}}\psi(t-\tau_{m})  \delta^{II}_{i} Q_{mi}(t-\tau_m)\phi(m)dm \\
 + \int_{M_{B_M}}^{M_U}\psi(t-\tau_m)  \delta^{II}_{i} Q_{mi}(t-\tau_m) \phi(m)dm    \\ 
\end{multline}

This equation is the same of Eq.(\ref{R(t)}) with the addition of $\delta^{AGB}_i, \delta^{Ia}_i, \delta^{II}_i$: these terms are the so called dust condensation efficiencies and represent the fractions of an element $i$ expelled from AGB stars, Type Ia and II SNe respectively, which goes into the dust phase of the ISM. In Eq.(\ref{R(t)dust}) we divided the third term of Eq.(\ref{R(t)}) into two integrals in order to separate the contribution between massive and low mass stars.

D98 and C08, in their works adopted arbitrary values for $\delta_i$, based on simple assumptions: for low and intermediate mass stars ($0.8-8 M_{\odot}$), $\delta_i^{AGB}$ were assumed equal to 1: it means that all the amount of carbon (C) or \textit{silicate-elements} (O,Mg, Si, Ca, and Fe) produced by a single star condensates into dust phase when the C/O ratio in the star ejecta is higher or lower than 1, respectively. For Type II and Type Ia SNe, $\delta_i^{II}$ and $\delta_i^{Ia}$ were set equal to 0.8 for both carbon and silicates.

In this work we adopt more recent and improved $\delta_i$ calculated by Piovan et al. (2011) (hereafter P11) for AGB stars and Type II SNe: these values depend on the mass and the metallicity of progenitor stars and have been derived from the comparison between theoretical studies and observational data. 
In the next paragraphs we will discuss the yields of dust by AGB stars and Type II SNe adopting these prescriptions and compare them with others widely employed in literature.

A separate discussion must be reserved to Type Ia SNe for which observational data and theoretical studies progressively changed their role concerning dust production.
During the last decade, the search for newly-formed dust in Type Ia SN remnants has been performed by Spitzer and Herschel satellites: most of the results attribute the IR emission to the shocked interstellar dust, with no detection of the newly formed one (Blair et al. 2007; Williams et al. 2012; Gomez et al. 2012). 
From a theoretical point of view, Nozawa et al. (2011) predicted that
dust formed in Type Ia SN explosions is destroyed before it can be injected into the ISM.  
The condensation efficiency $\delta_i^{Ia}$   was decreased by a factor 10 Pipino et al. (2011) with respect to D98 and C08. 
Here, as starting hypothesis, we set to zero the condensation efficiencies for each element as far as the contribution of Type Ia SNe is concerned ($\delta^{Ia}_i=0$). 
A possible contribution of dust production by Type Ia SNe is explored in Section~\ref{discussion}.
%%%%%%%%%%%%%%%%%%%%%%%%%%%%%%%%%%%%%%%%%%%%%%
\begin{figure}
\centering
\includegraphics[width=.47\textwidth]{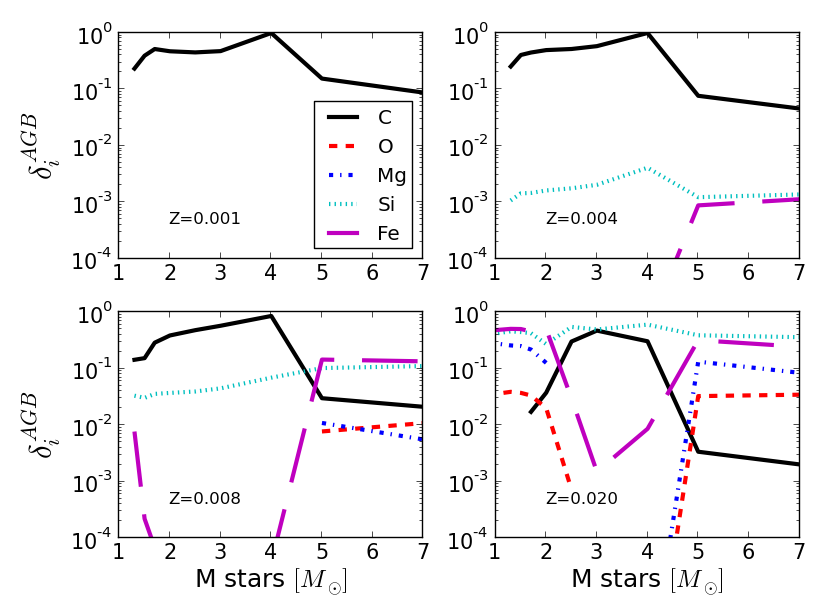}
\caption{Condensation efficiencies of C, O, Mg, Si and Fe for AGB stars as reported in Piovan et al. (2011) for different metallicities.
Black solid line for C, red dashed line for O, blue dash-dot line for Mg, cyan dotted line for Si and magenta long-dashed line for Fe. 
When condensation efficiencies are equal to zero, no lines are shown.}
\label{fig:Cond_eff}
\end{figure}
%%%%%%%%%%%%%%%%%%%%%%%%%%%%%%%%%%%%%%%%%%%%%%%%%
%%%%%%%%%%%%%%%%%%%%%%%%%%%%%%%%%%%%%%%%%%%%%%%%
\begin{figure}
\centering
\includegraphics[width=.47\textwidth]{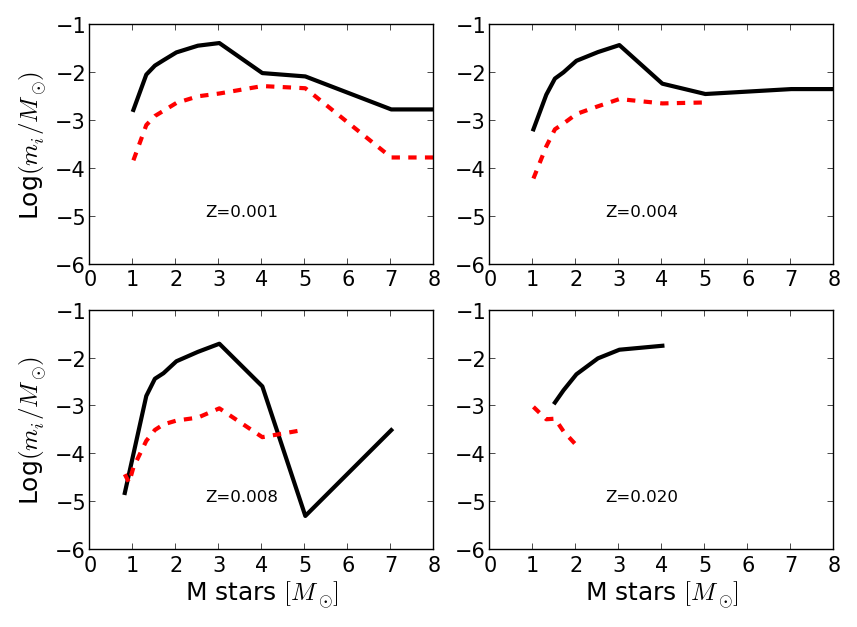}
\caption{Newly produced carbon (black solid lines) and oxygen (red dashed lines) mass of dust for AGB stars as a function of the initial stellar mass for four different metallicities. The dust yields have been reproduced taking into account condensation efficiencies by Piovan et al. (2011) and stellar yields from van den Hoek $\&$ Groenewegen (1997).}
\label{fig:dustAGBy}
\end{figure}
%%%%%%%%%%%%%%%%%%%%%%%%%%%%%%%%%%%%%%%%%%%%%%%%%
\subsubsection{Dust from AGB stars}
The cold envelope of AGB stars is a good environment in which nucleation and the formation of the first dust-seeds can occur. The total amount of dust produced in the previous phases of these stars is negligible because of the low amount of material in their ejecta and because the physical conditions of their winds do not favor its formation (Gail 2009). The dust species formed during the AGB-phase of low and intermediate mass stars (LIMS) strongly depend on their surface composition (Ferrarotti $\&$ Gail 2006). The stellar mass and metallicity play a key role in the number of thermal pulses that occur in pre-AGB phases and determine the surface composition, and therefore, the formation of particular dust species (Ferrarotti $\&$ Gail 2006; Nanni et al. 2013; Ventura et al. 2012). 

In Fig.~\ref{fig:Cond_eff} we show the condensation efficiencies of AGB stars as predicted in P11 for carbon, silicon, magnesium, oxygen and iron, which are the elements we focus in this work. $\delta_i^{AGB}$ values depend on the stellar mass and the metallicity, thus accounting for the dependence of the C/O ratio in their surfaces. $\delta_C^{AGB}$ dominates at lower metallicities and then decreases towards higher ones favoring the condensation of heavier elements. This is in agreement with AGB models of Nanni et al. (2013), where C/O in the stellar surfaces decreases with the metallicity.
In Fig.~\ref{fig:dustAGBy} we show the carbon and oxygen dust mass produced by AGB stars in the mass range $1-8 M_{\odot}$: the yields peak at mass values between 2 and  $3M_{\odot}$. Carbon is the only element with a non-negligible dust yield for two main reasons: it has a relatively high condensation efficiency and AGB stars are strong carbon producers. Carbon dust yields are comparable with the D98, Zhukovska et al. (2008) and Valiante et al. (2009) prescriptions.
On the other hand, elements heavier than carbon have lower condensation efficiencies and yields (Romano et al. 2010), which cause the net separation between the carbon and silicate dust production: here, it seems that AGB stars are not able to form silicates. Actually, LIMS can also produce some amount of such dust species: the yields presented in Fig.~\ref{fig:dustAGBy} reflect only the newly produced element abundance. In our chemical evolution model, it is also considered the fraction of the initial stellar mass that has not been processed in the inner region of the star and that during AGB-phases is expelled into the ISM: this returned mass composition reflects that of the star when it was formed, and could be chemically enriched by heavy elements. According to the differential condensation efficiencies, part of this mass can condensate, forming silicates in a non negligible mass fraction.

\subsubsection{Dust from Type II SNe}
Type II SNe are believed to cover an important role in dust production as witnessed by observations. Thanks to infrared and sub-millimeter studies, evidence of the presence of dust in the environment of historical supernova remnants like SN1987A (Danziger et al. 1991), Cas A and the Crab Nebula has been observed (Gomez 2013 and reference therein). Core collapse SNe are prolific dust factories, forming a total amount of dust between $0.1 M_{\odot}$ and $0.7 M_{\odot}$. Furthermore, the origin of dust in QSO hosts at high redshift, can be only explained by a particular source able to reproduce a consistent amount of dust mass in less than 1 Gyr of cosmic evolution. Stars with masses higher than 8$M_{\odot}$ can explode as Type II SNe in less than 30 Million years (Matteucci $\&$ Greggio 1986): for this reason they are believed to be the main source of dust in early epochs of galactic evolution (e.g. Maiolino et al. 2006), although some other studies consider that either AGB stars (Valiante et al. 2009, 
Dwek et al. 2011) or dust accretion (Pipino et al. 2011; Calura et al. 2014; Mancini et al. 2015) might play a significant role in the dust enrichment of such high redshift objects. 
Even if Herschel, SCUBA and more recently ALMA are acquiring a great deal of data, it is not easy to give a satisfactory estimate of the total amount of dust produced from Type II SNe. In particular, it is unknown what is the real effect of the reverse shock which has a typical time scale of $10^3-10^4 \textrm{yr}$, longer than the actual lifetime of historical SNe (Bianchi \& Schneider 2007).

\begin{figure}
\centering
\includegraphics[width=.47\textwidth]{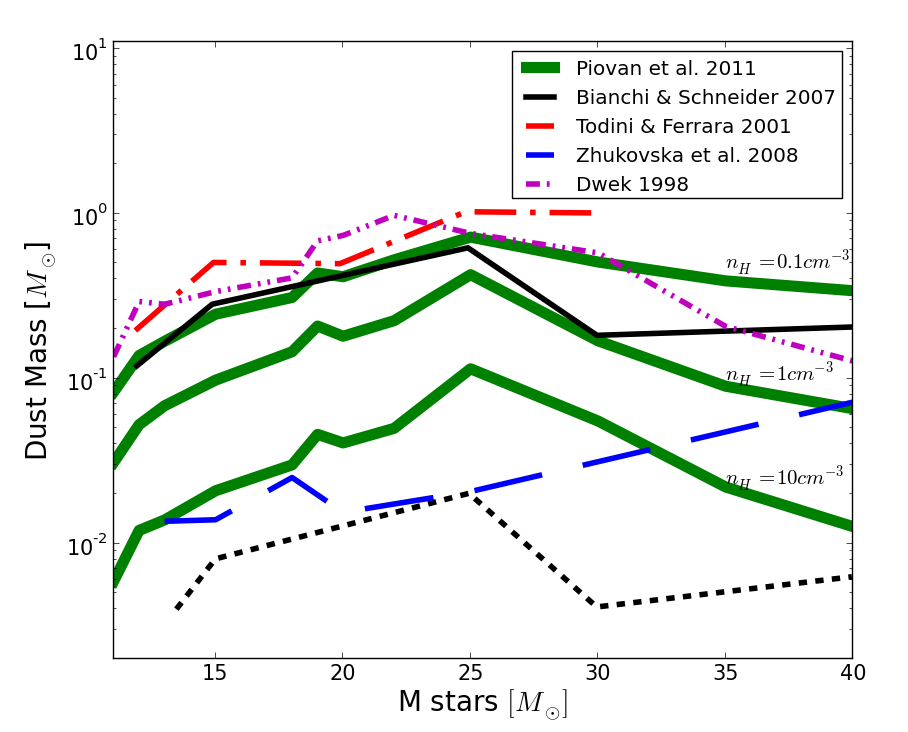}
\caption{Total dust amount [$M_{\odot}$] produced by Type II SNe as function of the initial stellar mass for different prescriptions. 
In green thick lines are presented the yields using $\delta_i^{II}$ of Piovan et al. (2011) for three different hydrogen densities: as the ambient density increase, the destruction process becomes more efficient, leading to lower dust yields.
In our work we adopt the yields which correspond to $n_{H}=1cm^{-3}$.
Todini $\&$ Ferrara (2001) prescriptions are shown in dash-dot red line, Bianchi $\&$ Schneider (2007) with and without considering the presence of the reverse shock in black solid and short-dashed line respectively, Zhukovska et al. (2008) with the contribution of all dust species in long-dashed blue line and Dwek (1998) in magenta dash-dot-dot line.}
\label{fig:dustSNy}
\end{figure}

In Fig.~\ref{fig:dustSNy}, dust yields from various authors are compared with the ones used in this work. Bianchi $\&$ Schneider (2007) revisited the previous work of Todini $\&$ Ferrara (2001) on Type II SNe dust yields predicting the formation of $0.1-0.6 M_{\odot}$ of dust in the ejecta of $12-40 M_{\odot}$ stellar progenitors. Considering also the presence of reverse shock, they concluded that only between $2$ and $20 \%$ of the initial dust mass can survive. In an independent work, Zhukovska et al. (2008) predict a similar amount of dust for such objects: their prescriptions come both from theoretical studies and observational data which included IR dust emission in SNRs, the amount of dust in historical SNe and studies on pre-solar grains in meteorites. Another important role is played by the environment surrounding the explosion of Type II SNe: the higher the density, the more resistance the shock will encounter and the more dust will be destroyed (P11, Nozawa et al. 2007). On the other hand, in a 
lower density environment, dust can easily resist to the passage of the shock, causing a more efficient dust formation.

Also for Type II SNe we adopt the dust $\delta_i{II}$ predicted by P11, which are calculated by taking into account the above-mentioned studies on dust formation, destruction as well as the density of the environment in which the SNe explode. Black lines in Fig.~\ref{fig:dustSNy} show the total dust production considering three different conditions for the density of neutral hydrogen, i.e. $\textrm{n}_{H}=0.1-1-10\,\textrm{cm}^{-3}$. In low density environments the amount of dust produced is similar to the prescription of Todini (2001) and Bianchi (2007), whereas in high density regions dust production becomes similar to the models where the reverse shock is included. In this work we adopt the yields of P11 corresponding to a neutral hydrogen density of $\textrm{n}_{H}=1\,\textrm{cm}^{-3}$: with this selection we do not overestimate nor underestimate previous prescriptions for high mass progenitors.
%%%%%%%%%%%%%%%%%%%%%%%%%%%%%%%%%%%%%%%%%%%%%%%%%%%%%%%%%%%%%%
%%%%%%%%%%%%%%%%%%%%%%%%%%%%%%%%%%%%%%%%%%%%%%%%%%%%%%%%%%%%%%%%%%%
%%%%%%%%%%%%%%%%%%%%%%%%%%%%%%%%%%%%%%%%%%%%%%%%%%%%%%%%%%%%

\subsection{Dust destruction} \label{destruction}
In literature it is possible to find various prescription describing dust destruction in the ISM. The main process for dust destruction is the sputtering in the ISM in high velocity SN shocks. The timescale of dust destruction is independent on the dust mass and it can be expressed as reported by C08:
\begin{equation} \label{time-scale-destruction}
T_{destr}=\dfrac{M_{ISM}}{(\epsilon \cdot M_{Swept})SN_{rate}}= \dfrac{M_{ISM}}{1360\cdot SN_{rate}}
\end{equation}
where ${M_{ISM}}$ is the mass of the ISM in the galaxy, $SN_{rate}$ represents the supernova rate and $M_{Swept}$ is the amount of mass swept up by the remnant. McKee et al. (1989) suggested $M_{Swept}=6800 M_{\odot} $ with an efficiency $\epsilon= 0.2$.
Zhukovska et al. (2008) predicted that the position of SNe in the galaxy influences the swept up mass. Furthermore, they adopted differential destruction time-scales depending on the dust species. Mancini et al. (2015) used prescriptions of de Bennassuti et al. (2014), where different parameters are adopted for core-collapse SNe and Pair-Instability ones ($M>100M_{\odot}$).

Another interesting study was made by Asano et al. (2013) (Hereafter A13), in which some differences can be highlighted with respect to Eq.(\ref{time-scale-destruction}). They suggest an efficiency $\epsilon=0.1$ and predict that $M_{swept}$ becomes smaller either when the density of the ISM is higher (because the amount of material that blocks SN blast is larger) or when the metallicity increases (the line cooling of metals is more efficient, leading to a lower temperature and higher density). Their prescription gives:
\begin{equation}\label{sweptup}
 M_{Swept}=1535\cdot n^{-0.202}\cdot[Z/Z_{\odot}+0.039]^{-0.289} [M_{\odot}]
\end{equation}
where $n=1.0\,cm^{-3}$ is the ISM density surrounding the SN environment. The swept mass in this case, assuming $Z/Z_{\odot}<1$ is always above the value of 1300$M_{\odot}$, as used in C08.

In this work we will use the updated metallicity-dependent prescriptions written in Eq.(\ref{sweptup}), whereas the destruction time-scale in Eq.(\ref{time-scale-destruction}) will be only used for comparison to the C08 model.
%%%%%%%%%%%%%%%%%%%%%%%%%%%%%%%%%%%%%%%%%%%%%%%%%%
%%%%%%%%%%%%%%%%%%%%%%%%%%%%%%%%%%%%%%%%%%%%%%%%%%

\subsection{Dust accretion}\label{accretion}
Some processes, like coagulation, increase the dust size favoring the formation of larger grain particles. As already pointed out, in this work we only follow the mass evolution of the dust and therefore we only consider as dust accretion the condensation of metals onto the surface of pre-existing dust grains. This process takes place efficiently in cold dense regions and, for this reason, it preferably occurs in molecular clouds rather than in the diffuse ISM. Since the pioneering work of D98, it was pointed out that grain growth is one of the fundamental ingredients in studying the dust mass evolution and, more recently, other studies support this thesis (A13, Hirashita et al.  2013, Valiante et al. 2011, Mancini et al. 2015).

In C08 the time-scale for the accretion is expressed as:
\begin{equation}
 \tau_{C08} = \tau_{0,i}/(1-f_i)
 \label{cagrowth}
\end{equation}
where $f_i$ represents the ratio between the dust and the gas phase of the element $i$. $\tau_{0,i}$ represents the typical lifetime of a molecular cloud which in C08 was kept constant for all elements at the value of $5\times10^7$ yr.

Dust accretion depends on many other parameters such as the fraction of molecular clouds in the whole galaxy rather than their fraction of metals. For this reasons Hirashita (2000) expressed the rate of dust accretion as follows:
\begin{equation}\label{higrowth}
\left[\dfrac{dM_{dust}}{dt}\right]=\dfrac{M_{dust} X_{cl} \, \chi_i}{\tau_{growth}}=\dfrac{M_{dust}}{\tau_{acc}}
 \end{equation}
where, $\chi_i=(1-f_i)$, $X_{cl}$ represents the fraction of the cool component in the ISM, $\tau_g$ is the characteristic dust growth time-scale and 
\begin{equation}
\tau_{acc}=\tau_g/(X_{cl}\chi_i)
\end{equation}
 Even if some elements (refractories) tend to be more depleted in dust grains with respect to others (volatiles), the dust composition consists substantially of metals. This 
 justifies the relation given by A13 between the growth time-scale and the metallicity:
 \begin{multline}\label{Asano_accr}
 \tau_{g}= 2.0 \times 10^7 yr \\ 
 \hspace{0.5cm} \times \left[ \dfrac{a}{0.1\mu m}\cdot \left(\dfrac{n_H}{100cm^{-3}}\right)^{-1} \cdot \left(\dfrac{T}{50K}\right)^{-\dfrac{1}{2}} \cdot \left(\dfrac{Z}{0.02}\right)^{-1} \right]
\end{multline}
 where the parameters reproduce the physical structure of a molecular clouds as well properties of dust grains. The typical time-scale $\tau_g=2\,Myr$ is reached when physical parameters are set on $T=50 K$ for temperature, $n=100 cm^{-3}$ for density and $\Bar a=0.1 \mu m $ for the typical size of grains.
 
In this work we use the prescription of Eq.(\ref{higrowth}) with the inclusion of the metallicity-dependent time-scale of Eq.(\ref{Asano_accr}).

\section{Model results}\label{dust_model_comparison}
\begin{figure}
\centering
\includegraphics[width=.47\textwidth]{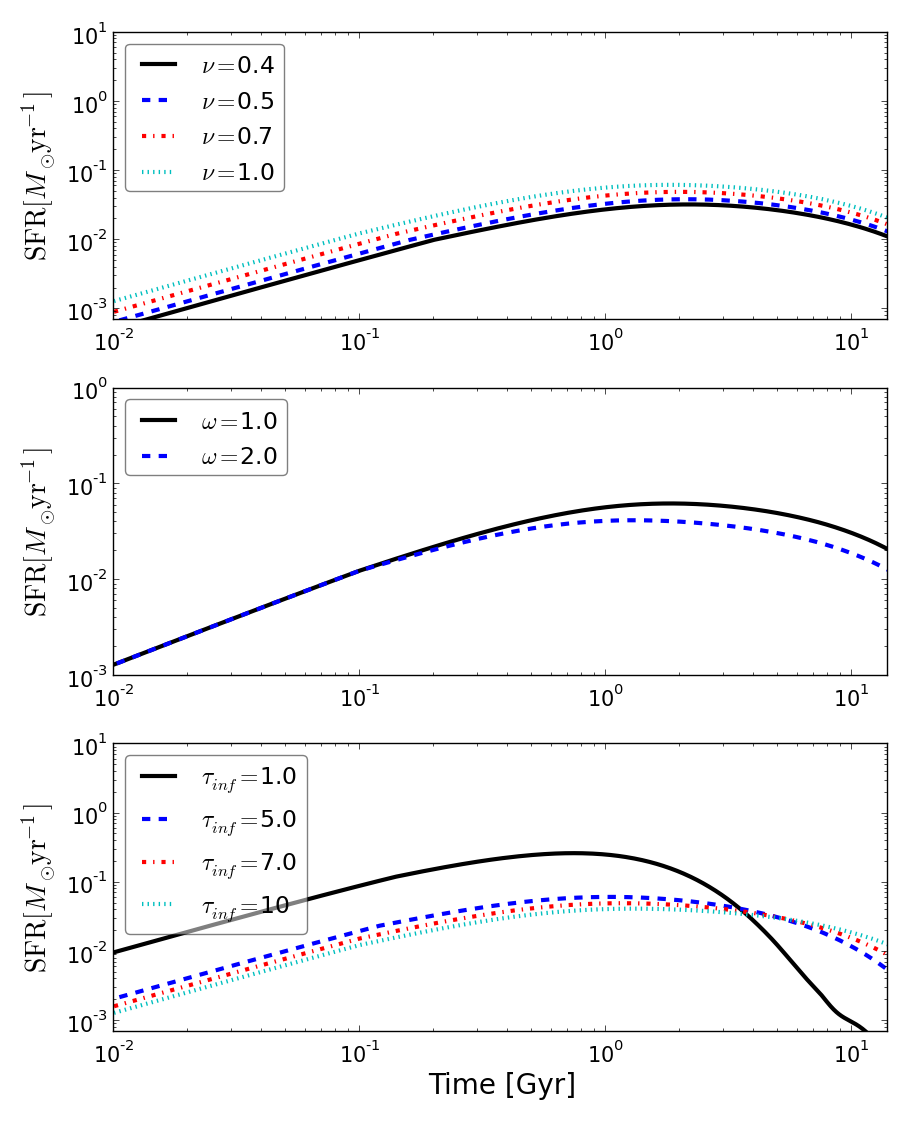}
\caption{Time evolution of the star formation rate for different values of some key parameters: in the top panel we varied $\nu$ from 0.4 to 1.0 $\textrm{Gyr}^{-1}$, in the middle panel $\omega$ from 1.0 to 2.0 and in the lower panel the $\tau_{infall}$ from 1.0 to 10 $\textrm{Gyr}^{-1}$.}
\label{fig:SFR}
\end{figure}

\begin{figure}
\centering
\includegraphics[width=.47\textwidth]{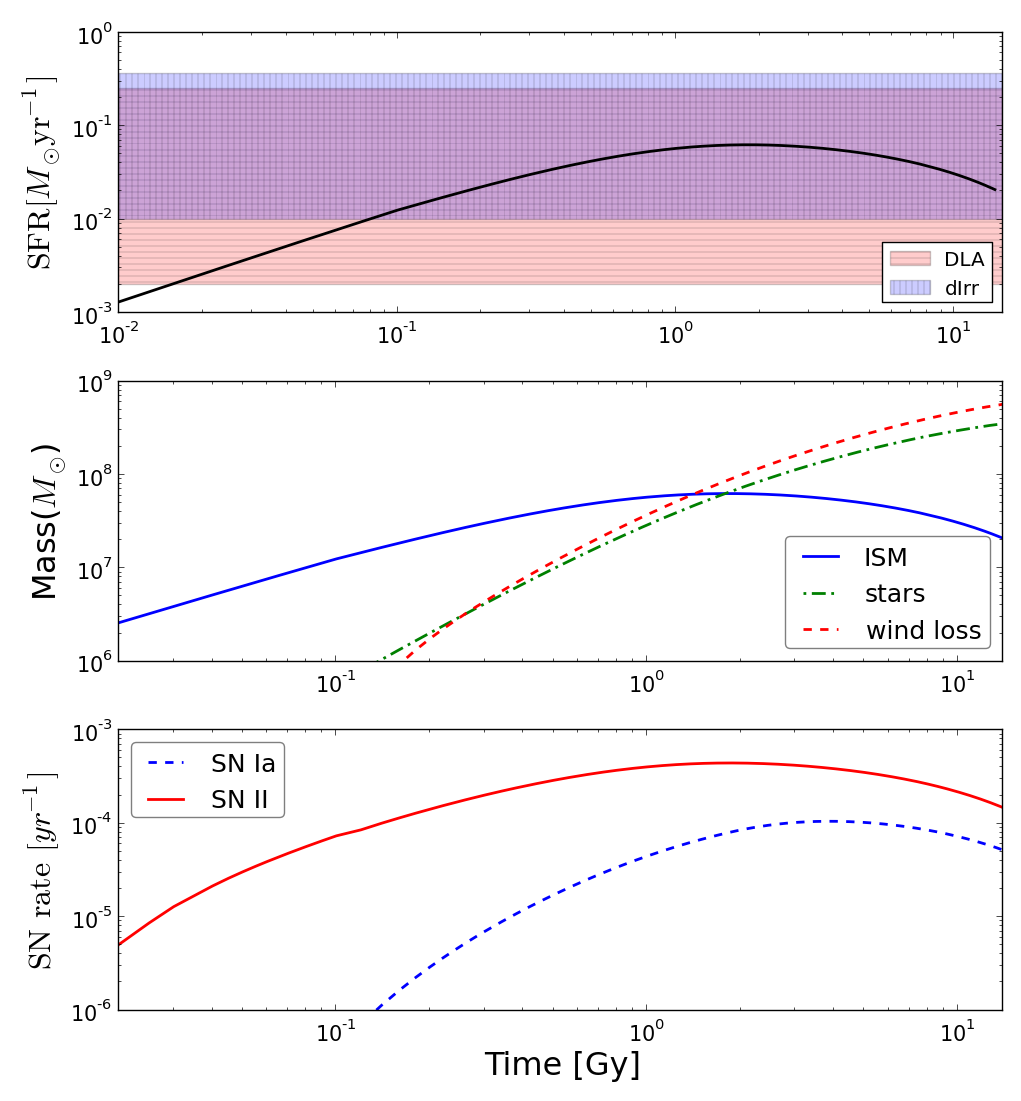}
\caption{Time evolution of model I3. In the top panel we show the evolution of the SFR in $M_{\odot} yr^{-1}$: 
the red shaded area represents the range estimated by Dessauges-Zavasdky et al. (2007) computed assuming a scale radius of 1 kpc as a typical size for DLAs, whereas the blue area represents the range measured in dwarf irregulars by Kennicutt et al. (2011).
In the middle panel we show the evolution of the mass (in solar masses) of the ISM (blue solid), stars (green dot-dashed) and the mass loss by galactic wind (red dashed). In the bottom panel we show the Type II and Type Ia SN rates (in number per year) in red solid and blue dashed, respectively.}
\label{fig:ISM}
\end{figure}

In this section we present the predictions for the evolution of dust in dwarf irregular galaxies for a set of chosen models. Dwarf irregulars are low mass and low luminosity galaxies which are characterized by on-going star formation. Many previous works have constrained the parameters of chemical evolution models for these galaxies. 
They also showed the effects of varying the model parameters, for irregulars but also for galaxies of different morphological type: 
Bradamante et al. (1998) varied the star formation efficiency, IMF and SFH, Romano et al. (2005) - Molla (2015) the IMF, Calura et al. (2009) the star formation efficiency, Romano et al. (2010) the stellar yields. 
Recently C\^ot\'e et al. (2016) studied the uncertainties in galactic chemical evolution models for the Milky Way, but without making a direct comparison with observational data. 
This paper is not focused on the precise characterization of a peculiar irregular galaxy: here, we use instead a typical dwarf irregular galaxy model, 
which is characterized by a low and continuous star formation, a relatively long infall time-scale and a moderate galactic wind rate, 
as already adopted in previous works, in particular in Calura et al. (2008).
The various models we considered and the prescriptions adopted for dust processing are reported in Table~\ref{dust_param}. In this Table we show the model name in the first column, we report the assumed star formation efficiency in the second column, in the third column the infalling mass, in the fourth column the time-scale of the infall, in the fifth column the wind efficiency, in the sixth column the assumed IMF and in the seventh column the references to the assumed dust condensation efficiencies are reported. Finally, in column eight and nine we indicate the references to the assumed dust destruction and accretion time-scales, respectively. 
Our typical irregular galaxy model is represented by the I0, I1 and I3 models, which only differ for dust prescriptions. 
In fact, we are particularly interested in the variation between the different dust prescriptions: in particular, model I3 represents our best model, as it includes the most updated ones. 
Here, we also performed some tests by changing the model parameters: in more detail, we found that acceptable values for the star formation efficiency, $\nu$, lie in the range 0.4-1.0 $\textrm{Gyr}^{-1}$; the time scale of the infall has a modest effect on the abundances and it lies in the range 5-10 Gyr. Finally, the wind parameter can vary between 1 and 2.5 in order to obtain acceptable results concerning the abundance patterns. 
In Fig.~\ref{fig:SFR} we present the effects on the SFR by varying such parameters. 
In the top panel we show that the SFR increases with $\nu$, although a very small difference is found between models with $\nu=1$ and $\nu=0.4$. 
In the middle panel we show that the spread caused by the wind parameter is also very small and it appears important only at late epochs, when the galactic wind is active.
In the bottom panel of the same Figure, one can see that the variation of the SFR is negligible when the time of the infall is larger than 5.0 Gyr. 
We obtain different results only when the time-scale of the infall is lower (i.e. 1.0 Gyr): the SFR reaches higher values at early epochs, because of the high amount of gas available to form stars, whereas at the end of the simulation is lower, because all the gas has been already consumed. 
The spread shown in Fig.~\ref{fig:SFR} is in agreement with the typical behaviour of irregulars, which, on average, are characterized by a smooth SFR with typical variations of factors 2-3 (Grebel 2004) and present time values between 0.001 and 0.36 $M_{\odot} yr^{-1}$ according to their stellar mass (Kennicutt et al. 2011). 
For the rest of this paper model I3 will be regarded as the fiducial one, 
as it accounts for the star formation rate values measured in DLAs as well as for their observed abundance pattern. 
In Fig.~\ref{fig:ISM} we show the evolution of some quantities related to model I3. In the top panel we show the behaviour of the SFR in comparison with the values measured in irregulars and DLA systems. 
In the middle panel, we show the time evolution of the masses of stars, ISM and the mass lost by galactic wind: at the beginning, the mass of the ISM represents the major component of the galaxy and it increases as the pristine infalling gas accretes to form the galaxy. As the SFR rises, also the stellar mass and the SN rate increase, as visible in the middle and bottom panel, respectively. 
This causes the formation of a galactic wind which considerably reduces the mass of the ISM. In the bottom panel of Fig.~\ref{fig:ISM}, we show the predicted Type II and Type Ia supernova rates, which are important both for dust and gas components. The trend of Type II SN rate traces the SFR one, because of the very short typical time-scales involved (from 1 to 30 Myr). On the other hand, Type Ia SNe are characterized by longer time-scales (from 30 Myr up to the time of the Universe). 
Finally, we tested the effect of suppressing the chemical enrichment from massive stars ($M>18M_{\odot}$) by assuming that these stars implode as black holes instead of exploding as SNe.
As this is the first time that this issue is included in a chemical evolution model with dust, we will show and discuss the results of this test in a dedicated paragraph (\ref{cut-off}).

%%%%%%%%%%%%%%%%%%%%%%%%%%%
%%%%%%%%%%%%%%%%%%%%%%%%%%%
\begin{table*}
\begin{center}
\begin{tabular}{c c c c c c | c c c}
\hline
 name & $\nu [Gyr^{-1}]$& $M_{infall} [M_{\odot}]$ & $T_{infall} [Gyr]$& $\omega$ & IMF & $\delta_{i}$ & $T_{destruction}$ & $T_{accretion}$ \\
\hline
\multicolumn{9}{c}{Dwarf irregular galaxy model}\\
\hline
 I0	&  1.00	& $10^9$ 	      & 10	&	1.00  &	Salpeter 	& D98	&	D98		   & 	---	 \\ 
 I1	&  1.00	& $10^9$ 	      & 10	&	1.00  &	Salpeter	& P11	&	A13		   & 	---	 \\
 I2	&  1.00	& $10^9$ 	      & 10	&	1.00  &	Salpeter	& D98	&	D98		   & 	D98	 \\
 I3	&  1.00	& $10^9$ 	      & 10	&	1.00  &	Salpeter	& P11	&	A13		   & 	A13	 \\
 I4	&  1.00	& $10^9$ 	      & 10	&	6.50  &	Salpeter	& P11	&	A13		   & 	A13	 \\
 \hline 
 \end{tabular}
 \caption{Parameters adopted for the different chemical evolution models. In columns from left to right we show the name of the model, the star formation efficiency (expressed in $Gyr^{-1}$), the mass of the galaxy (in $M_{\odot}$), the infall time-scale (in $Gyr$), the wind parameter (dimensionless) and the adopted IMF. In the seventh, eighth and ninth columns we show the different prescriptions used for condensation efficiencies ($\delta_i$), destruction and accretion time-scales, respectively. D98, P11 and A13 refer to prescriptions of Dwek (1998), Piovan et al. (2011) and Asano et al. (2013), respectively.}
 \label{dust_param}
\end{center}
\end{table*}

%%%%%%%%%%%%%%%%%%%%%%%%%%%%%%%%%%%%%%%%%%%%%%%%%%%
%%%%%%%%%%%%%%%%%%%%%%%%%%%%%%%%%%%%%%%%%%%%%%%%%%%
%%%%%%%%%%%%%%%%%%%%%%%%%%%%%%%%%%%%%%%%%%%%%%%%%%%
%%%%%%%%%%%%%%%%%%%%%%%%%%%%%%%%%%%%%%%%%%%%%%%%%%%
%%%%%%%%%%%%%%%%%%%%%%%%%%%%%%%%%%%%%%%%%%%%%%%%%

\subsection{Dust to gas ratio}
Here we present the dust-to-gas ratio predicted  by our models for a typical irregular galaxy undergoing different possible types of dust evolution. In practice, we changed the parameters 
related to dust processing (see Table \ref{dust_param}) while keeping fixed the remaining parameters concerning the evolution of the ISM.

\begin{figure}
\centering
\includegraphics[width=.47\textwidth]{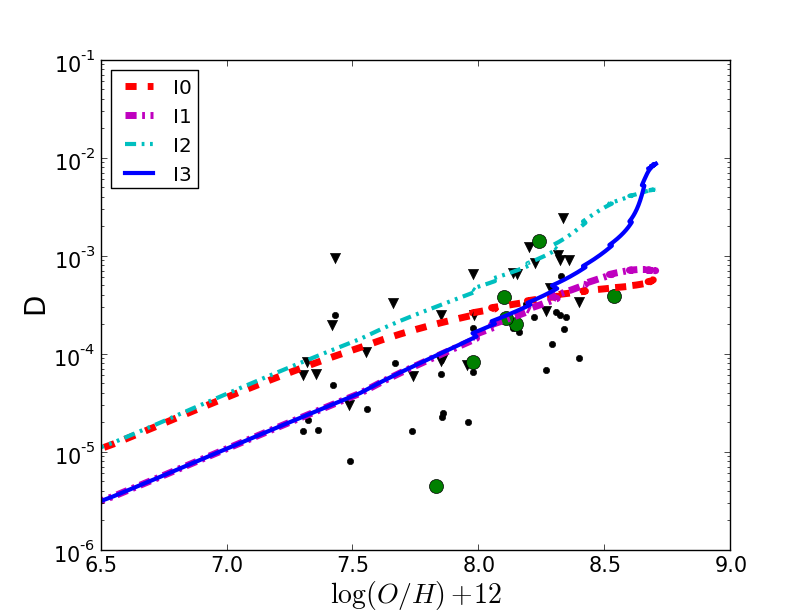}
\caption{Dust-to-gas ratio versus metallicity.
\textit{Models}: Dashed red line refers to I0 model, magenta dot-dashed line to I1, cyan dot-dot-dashed line to I2 and blue solid line represents model I3.
\textit{Data}: black dots and triangles represent values and upper limits respectively for local dwarf irregulars by Lisenfeld $\&$ Ferrara (1998) and big green circles from Hirashita et al. (2008).}
\label{fig:datacomp}
\end{figure}

In Fig.~\ref{fig:datacomp} we show the ratio between the mass of dust and neutral hydrogen, $D=M_{dust}/M_{HI}$, versus the metallicity for different chemical evolution models and observations. Data were taken from Lisenfeld $\&$ Ferrara (1998) (local dwarf irregulars), Hirashita et al. (2008) (blue compact dwarf spheroidals), which have been collected in Galametz et al. (2011). Those authors estimated the amount of dust mass from far-infrared measurements and compared them with the amount of neutral hydrogen and oxygen. 

I0 and I1 models (red dashed and solid, respectively), only consider dust production and destruction: I0 model, which adopts D98 prescriptions, produces a higher amount of dust with respect to I1, which adopts the updated ones. 
The two curves differs at the lowest metallicities: the offset between the two is cause by the major amount of dust ejected by Type II and Ia SNe when D98 $\delta_i$ are adopted (see section~\ref{sec_composition}). For $\log (O/H)+12>8$, the two models converge and predict a same value for the dust-to-gas ratio.

Dwarf galaxies have very low star formation efficiency, which could reflect the paucity of molecular clouds present in these environments. In C08, dust accretion was not taken into account, as this process preferably occurs in such clouds. However, we tried to include dust accretion in other models because it seems to play a fundamental role during the entire evolution of a galaxy (see section~\ref{following_paragraph}, Asano et al. 2013, Dwek et al. 2011, Mancini et al. 2015). 
The I2 model contains the same prescriptions used in C08, including dust accretion: the total amount of dust in model I2 is higher than in I0 and the discrepancy increases at higher metallicities. The separation between the two models is evident since $\log (O/H)+12>7$ values, indicating that in I2 dust accretion becomes the dominant process since the early phases of dust evolution.
Finally, model I3 adopts P11 $\delta_i$ and the new prescriptions for the swept up mass and for the accretion time-scale, as described in Eqs. (\ref{sweptup}) and (\ref{Asano_accr}), respectively. In this case, the discrepancy from I1 is negligible for lower metallicities, becoming larger when $\log (O/H)+12>8$. It means that in I3 model, dust accretion becomes important at higher metallicities or, in other words, at longer evolution time-scales with respect to I2. 

I2 and I3 are the only models which can reproduce the observed high amount of dust. This indicates that dust accretion performs an important role in dust evolution and it should be modeled in a proper way: the I3 model better reproduces the trend observed in data, which seems to increase steeper with the metallicity.
%%%%%%%%%%%%%%%%%%%%%%%%%
%%%%%%%%%%%%%%%%%%%%%%%%%%%%%%%%%%%%%%%%%%%%%%%%%

\subsubsection{Massive star cut-off}\label{cut-off}
For a long time, the scientific community has reserved great interest to the study of massive stars (above $10M_{\odot}$). 
In spite of this, nowadays it is not completely clear which stars do explode as SNe and which ones do collapse to a black hole.
In fact, recent theoretical and observational studies suggest the possibility that some massive stars may directly collapse to a black hole without enriching chemically the ISM.
From a theoretical point of view, it has been known for a long time that very high mass star explosions are more unlikely than those of lower masses (Fryer 1999).
O'Connor $\&$ Ott (2011) concluded that stars heavier than $20M_{\odot}$ have difficulty in exploding. 
More recently, simulations by Ugliano et al. (2012) have shown that core-collapse explosion and direct black hole formation are both possible outcomes for progenitor stars above 15$M_{\odot}$.
From the observational point of view, there is no evidence for SN progenitor mass above $\sim20M_{\odot}$ (Smartt 2009).
Smartt (2015) explores the evolution of massive stars after the quiescent phase on the Main Sequence: the most probable scenario shows that stars above $18M_{\odot}$ fail SN explosion and directly collapse to a black hole.
However, this is a strong assumption and many uncertainties are still present. 
An indication of the fact that this problem is not completely understood comes from Ugliano et al. (2012): in fact, their predictions do not take into account binary effects and, despite all black holes should swallow the progenitor star, they found an exception for a mass progenitor star of 37 $M_{\odot}$. 
They also concluded that a direct comparison with observations requires caution.

Massive stars are very important actors in chemical evolution models as they are responsible for metal production. 
For this reason, model predictions can be affected when the chemical enrichment of massive stars is not considered. 
In addition, even if massive stars do not explode as SNe,  
the mass loss via stellar wind integrated over the stellar life,
contributes to the chemical enrichment of the ISM and cannot be neglected, especially for C and He. 
Brown $\&$ Woosley (2013) have already tested the effect of cutting chemical enrichment from massive stars in the chemical evolution of the solar neighborhood.
In their model, they cannot reproduce the observed abundances without the chemical contribution of stars above $18 M_{\odot}$.
On the other hand, they slightly reproduce the chemical pattern when the cut-off is moved to $25 M_{\odot}$, and even better to $40 M_{\odot}$. 

\begin{figure}
\centering
\includegraphics[width=.47\textwidth]{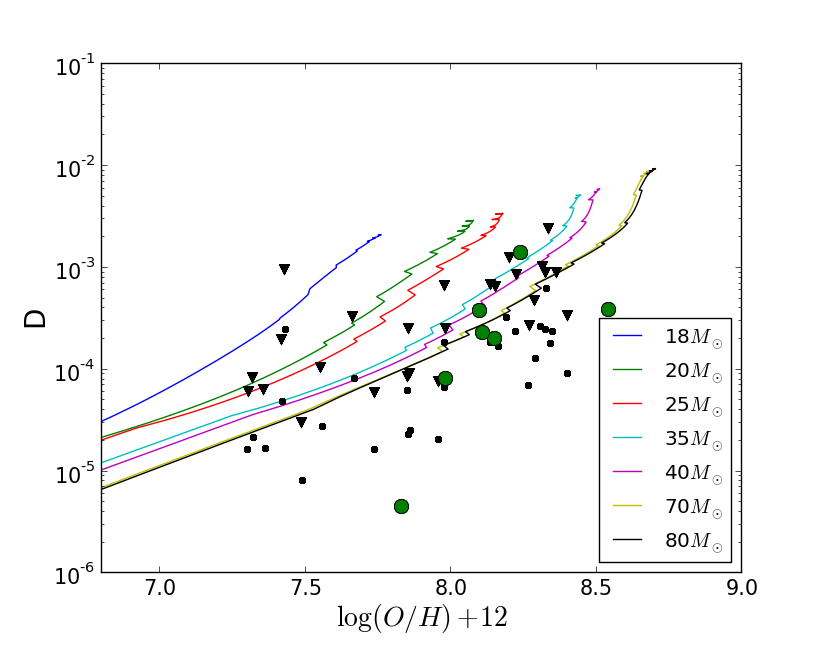}
\caption{$D=M_{dust}/M_{HI}$ versus metallicity as in Fig.~\ref{fig:datacomp}. 
The lines of various colours represent model I3 computed assuming different values for the cut-off mass (as reported in the insert). 
}
\label{fig:cut-off}
\end{figure}

In this article we test for the first time the effect of the mass cut-off concerning high mass stars in the chemical evolution model which takes into account the presence of dust. 
A mass cut-off of $18M_{\odot}$ means that the chemical contribution of the stars above this mass is not taken into account.
In Fig.~\ref{fig:cut-off} we show the evolution of model I3 with different cut-off masses.
We found disagreement between models and data when the cut-off mass is set below 25 $M_{\odot}$. 
On the other hand, our models reproduce the data when a mass cut-off $\ge 35 M_{\odot}$ is adopted:
in fact, above this value, different cut-off masses lead to a little scatter and all of the models are able to reproduce the bulk of the data.
Concerning the upper value of D of all different models we did not find big differences: 
this indicates that the dust production is not deeply affected by the cut-off mass.
It is not possible to say the same for the metallicity: in fact, massive stars are very important metal producers, especially for alpha elements such as oxygen.
It is impossible to reach $log(O/H)+12 > 7.7$ when we cut the contribution of massive stars above $18M_{\odot}$, 
whereas a cut off of 25 $M_{\odot}$ leads to $log(O/H)+12>8$ and is more acceptable. 
Finally, in order to reproduce the highest values of $log(O/H)$ showed by the data, exploding masses up to 35-40 $M_{\odot}$ are necessary.
In the rest of the paper we will not further consider this cut-off, as we are most interested in the evolution of the dust and the effect that it produces on the observed chemical abundances of the ISM.
%%%%%%%%%%%%%%%%%%%%%%%%%
%%%%%%%%%%%%%%%%%%%%%%%%%%%%%%%%%%%%%%%%%%%%%%%%%

\subsection{Dust composition}\label{sec_composition}
In this paragraph we focus on the composition of the dust. For the sake of comparison with D98 and C08 we consider two species of dust, namely silicates and carbonaceous dust. Following the definitions of D98 and C08, silicates are dust particles formed by O, Mg, Si, S, Ca and Fe elements, while C dust consists of carbonaceous solids.
The possible presence of a iron-rich dust different from silicates will be discussed in Section \ref{DLA_sec}.

\begin{figure}
\centering
\includegraphics[width=.47\textwidth]{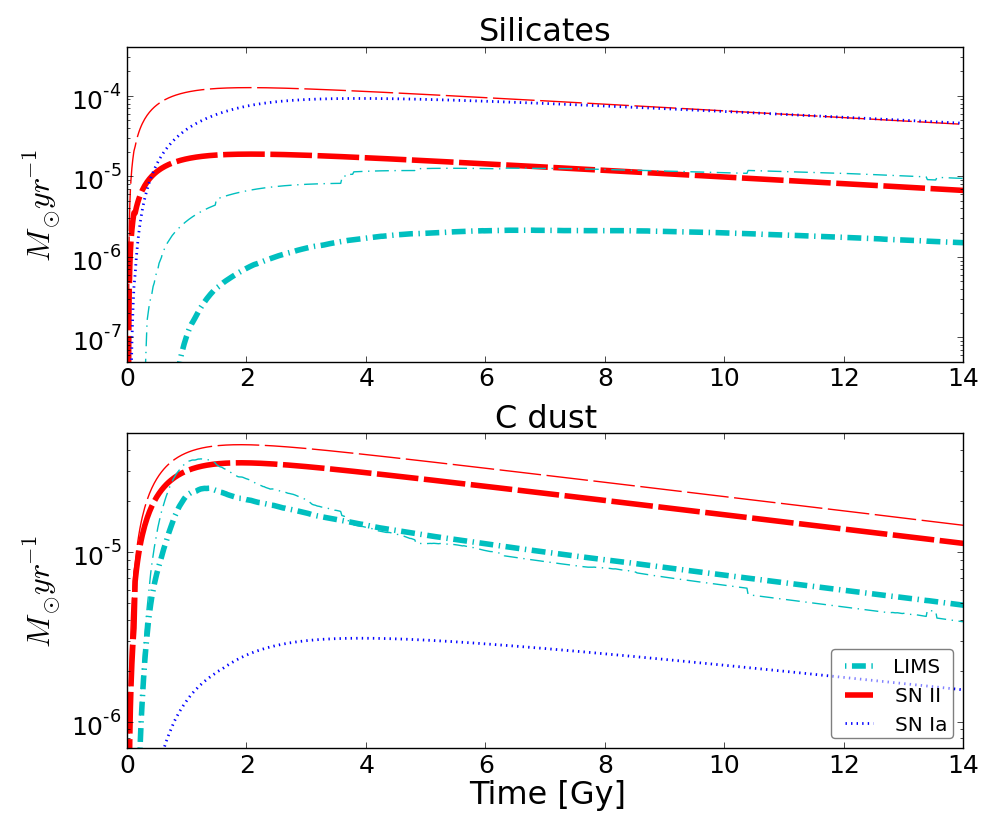}
\caption{Production rate of carbon and silicate dust in a dwarf irregular galaxy. Red dashed and cyan dot-dashed lines correspond to Type II SNe and LIMS dust sources respectively. Thick lines are for the results using P11 dust condensation efficiencies (model I1), while thin lines correspond to D98 ones (model I0). In blue dotted line we show the dust contribution of Type Ia SNe, as predicted in D98.}
\label{fig:dust_comp_cfr}
\end{figure}

In Fig.~\ref{fig:dust_comp_cfr} we show the formation rate of silicates and carbonaceous dust predicted using the dust condensation efficiencies of D98 (thin lines) and P11 (thick lines). In the bottom panel one can see how the two dominant carbon dust sources are the Type II SNe and low and intermediate-mass stars (LIMS). 
As Type II SNe have shorter lifetimes with respect to LIMS (Padovani $\&$ Matteucci 1993 and Matteucci $\&$ Greggio 1986), they dominate the dust production in the earliest epochs of the galactic evolution. AGB stars strongly contribute to pollute the ISM with carbonaceous dust and they become the major producers, remaining comparable with Type II SNe until the present time. Type Ia SNe do not influence the balance of carbonaceous grains.

The evolution of silicate dust production is presented in the top panel of Fig.~\ref{fig:dust_comp_cfr}: the bulk of this species is almost entirely produced by Type II SNe, whereas LIMS play a negligible role.

When D98 prescriptions are adopted, a higher dust production rate from Type II SNe and AGB stars is reached with respect to the model in which $\delta_i$ from P11 are considered. Furthermore, Type Ia SNe play a fundamental role, as they produce lots of iron that in this case is counted among silicates: actually, the iron from Type Ia SNe could be accreted in a separate, iron-rich dust species, as we discuss at the end of Section~\ref{DLA_sec}.
%%%%%%%%%%%%%%%%%%%%%%%%%%%%%%%
%%%%%%%%%%%%%%%%%%%%%%%%%%%%%%%
%%%%%%%%%%%%%%%%%%%%%%%%%%%%%%%
%%%%%%%%%%%%%%%%%%%%%%%%%%%%%%%
\subsection{Following dust evolution}\label{following_paragraph}
Depending on the SFH of a galaxy, the fraction of the dust and the contribution of the different processes occurring in the ISM can vary. In our model we can differentiate them and know how they evolve during the galactic time. 

\begin{figure}
\centering
\includegraphics[width=.47\textwidth]{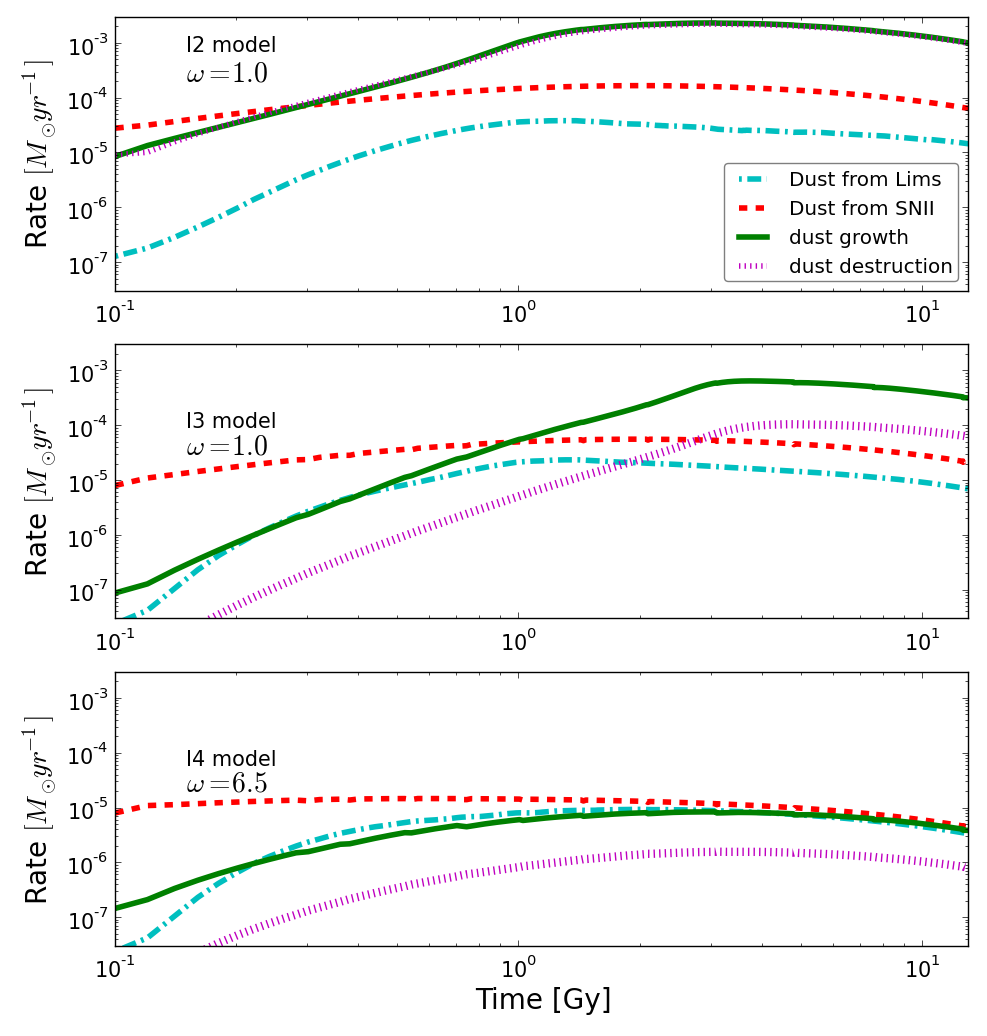}
\caption{Rate in solar masses per year of the different processes governing the dust evolution. The dashed-dot cyan and dashed red lines show the production rate by AGB stars and Type II SNe, respectively; the solid green line indicates the rate of the accretion process while the magenta dotted line represents the destruction rate. 
Predictions for I2, I3 and I4 models are shown in the top, middle and bottom panel respectively.}
\label{fig:following}
\end{figure}

In the top and middle panel of Fig.~\ref{fig:following} we compare the rates of accretion, destruction, and production of dust versus time for the I2 and I3 models, respectively. In this figure we are able to compare directly the different evolution of the dust when C08 or updated recipes are adopted. 
For the I2 model, dust destruction and accretion trace each other because of the definition of their typical time-scales are both proportional to the characteristic time of the star formation (see D98). Stars poorly contribute to the injection of dust in the ISM and the accretion process plays the major role during the whole galactic time.
In the middle panel the evolution of I3 model we show, and some differences emerge. In the initial phase, dust production by Type II SNe is the most important process: in fact, these massive stars ($8-40M_{\odot}$) have short lifetimes and rapidly inject dust into the ISM. According to Eq. (\ref{higrowth}), the accretion rate increases because the dust amount in the ISM becomes higher and the infalling gas continues to accrete. In addition, Eq. (\ref{Asano_accr}) shows that the accretion time-scale becomes shorter as the metallicity increases, strengthening the rise of the accretion rate. At a certain point, the rate of dust production and accretion become equal and then, dust accretion dominates until the end of the simulation. This result is in agreement with prediction of Asano et al. (2013) who defined the so called \textit{critical metallicity} as the metallicity at which the contribution of dust accretion overtakes the dust production from stars. In I3 model the critical metallicity assumes a 
value $Z_{crit,I3}=0.442Z_{\odot}$ ($Z_{\odot}=0.0134$, Asplund et al. 2009). As already mentioned, for I2 model the accretion rate becomes important at early epochs and we found a lower value for the critical metallicity, $Z_{crit,I2}=0.172Z_{\odot}$.

It is interesting that dust destruction plays a negligible role, while the most important process able to decrease the amount of dust is the galactic wind: in fact, it not only removes directly the dust from the galaxy, but also regulates the efficiency of the accretion process. To better fix this concept, we report in the bottom panel of Fig.~\ref{fig:following} the dust evolution for I4 model, characterized by a higher galactic wind parameter ($\omega=5$). In this case, dust accretion always lies below dust production and the critical metallicity is not reached: the galactic wind removes gas from the ISM which cannot condensate onto pre-existing dust grains any more, causing a reduction in the accretion rate. As before, in the earliest epochs the dust production rate is dominated by Type II SNe, but it decreases as soon as the galactic wind starts. On the other hand, the galactic wind does not deeply influences the dust production rate by 
AGB stars and, in this scenario, their contribution is similar to the one of Type II SNe. Galactic wind also decreases the destruction process: in such cases the supernova rate decreases with the star formation (see Fig.~\ref{fig:ISM}), leading to higher values of destruction time-scales, according to Eq. (\ref{sweptup}).
%%%%%%%%%%%%%%%%%%%%%%%%%%%%%%%
%%%%%%%%%%%%%%%%%%%%%%%%%%%%%%%
%%%%%%%%%%%%%%%%%%%%%%%%%%%%%%%
%%%%%%%%%%%%%%%%%%%%%%%%%%%%%%%

\section{Comparison with DLA systems} \label{DLA_sec}
In this section we compare our model with data of Damped Lyman Alpha systems (DLAs).
DLAs are a class of QSO absorbers, with neutral hydrogen column density $N(HI)\geq 10^{20.0} cm^{-2}$ and lying in the typical redshift range between 1 and 5 (Wolfe et al. 2005). They are the best observables available of the ISM in the high redshift Universe. Thanks to high resolution spectroscopy, it is possible to measure high precision column densities from their spectra. Observations show a variation in metallicity during the cosmic time, revealing that these systems can be seen in different stages of their evolution. For this reason DLA systems offer a great opportunity for studying the composition of the ISM at different cosmic epochs and evolutionary stages. 
The nature and the morphological type of DLA-host galaxies have been the subject of a long debate (see, e.g., Wolfe et al. 2005). The comparison with stellar abundances in local dwarf galaxies (Salvadori $\&$ Ferrara 2012; Cooke et al. 2015) or the study of $\alpha$-element abundances with chemical evolution models (Matteucci et al. 1997; Calura et al. 2003), are in agreement in associating DLA systems to dwarf star forming galaxies. Based on these previous results, we adopt our models of dwarf irregulars to test the behavior of dust in DLA systems. 
In the top panel of Fig.~\ref{fig:ISM} we show the model SFR compared with values measured in DLAs and in irregular galaxies.

\subsection{The method}
Comparing model results with observed abundances in DLA systems requires a specific methodology because the abundance measurements in such systems only refer to the gas phase of the ISM. This is because atoms or ions incorporated into dust grains cannot be detected with absorption-line spectroscopy and, as a result, the measured elemental abundances are depleted with respect to the total interstellar abundance (gas plus dust). The effects of dust in DLA systems have been considered in several papers (e.g., Junkkarinen et al. 2004; Vladilo et al. 2006; De Cia et al. 2013). Calura et al. (2003) performed depletion corrections on DLA data in order directly compare them with the total abundances predicted by chemical evolution models. 

Here we follow a different approach: 
with our model we  calculate the amount of dust in the ISM and, by tracking the dust production, destruction and accretion of the different elements, we calculate their fractions in the dust phase.  By subtracting this fraction from the total, we predict the gas-phase abundances that can be directly compared with the DLA abundance measurements. In this procedure, the condensation efficiencies of individual elements, $\delta_i$, as well as the prescriptions for the accretion and destruction described in Section~\ref{dust_model}, are used to calculate the individual gas-phase abundances. 
We focus on two refractory elements, iron and silicon, that are commonly measured in DLA systems and that are expected to be incorporated in dust form. Carbon abundances, which could be useful to test the presence of carbonaceous dust, are extremely rare in DLA systems due to the saturation of the interstellar carbon lines.

In order to tune the parameters of the DLA model  we also use volatile elements, such as zinc and sulfur, which are not expected to be incorporated into the dust.
Local interstellar observations (Jenkins 2009) and calculations of condensation temperatures (Lodders 2003) suggest that zinc and sulfur are mostly volatile in nature, even though the case of sulfur is not completely clear (Calura et al. 2009; Jenkins 2009). In high-density, molecular gas a fraction of zinc and sulfur might be incorporated in dust, but this is not a reason of concern in our case since the molecular fraction is generally very low in DLA systems (Ledoux et al. 2003).
Following the procedure of Vladilo et al. (2011), hereafter V11, we first tune the parameters of galactic chemical evolution using the S/Zn ratio, which is unaffected by dust depletion processes both in the model predictions and in the data.
We then use S and Zn as a reference to measure the relative abundances of the refractory elements Si and Fe, i.e. we study the ratios Si/S, Si/Zn, Fe/S, and Fe/Zn. 
The models predictions for element-to-element ratios of this type are more robust than the predictions for absolute abundances (relative to hydrogen). 
At the same time, these ratios are strong indicators of the possible presence of dust, since in each case they represent a ratio between a refractory element, affected by dust processes,
and a volatile element, not affected by dust processes.

We assume that the elemental abundances in the galactic ISM are determined by two processes: 1) chemical enrichment of the gas by stellar ejecta and 2) elemental depletion caused by the condensation of the gas onto dust particles. In principle, one should also take into account  the ionization state of the gas for a precise conversion of the column densities into abundances, but ionization corrections for DLA measurements are generally smaller than the column density errors (Vladilo 2001).

\subsection{DLA data}\label{sec_DLA_data}
The dataset used in this work is the same of V11, with the addition of 34 DLA systems with associated measurements collected in the last years.
We report in Table~\ref{tab:DLA_sample} all the data not present in V11: in particular, in Table~\ref{tab:DLA_sample} we report the name of QSO in column one and the absorption redshift in column two, while in the third, fourth, fifth and sixth column are presented the column density measurements of zinc, sulfur, iron and silicon, respectively. In the last column are reported the codes for each column density which are associated to the literature references listed in Table~\ref{tab:references}.
Before comparing the data with the model predictions we have lowered and increased the ZnII and SII column densities, respectively by 0.1 and 0.04 dex to take into account the recent redetermination of the relative oscillator
strengths provided by Kisielius et al. (2014; 2015). The database of DLA ZnII and SII column densities found in the literature is instead based on Morton's (2003) oscillator strengths. In Table~\ref{tab:DLA_sample} we give the original column densities, before the application of this correction.

\subsection{Results and comparison}\label{sec_DLAresults}
\begin{figure}
\centering
\includegraphics[width=.47\textwidth]{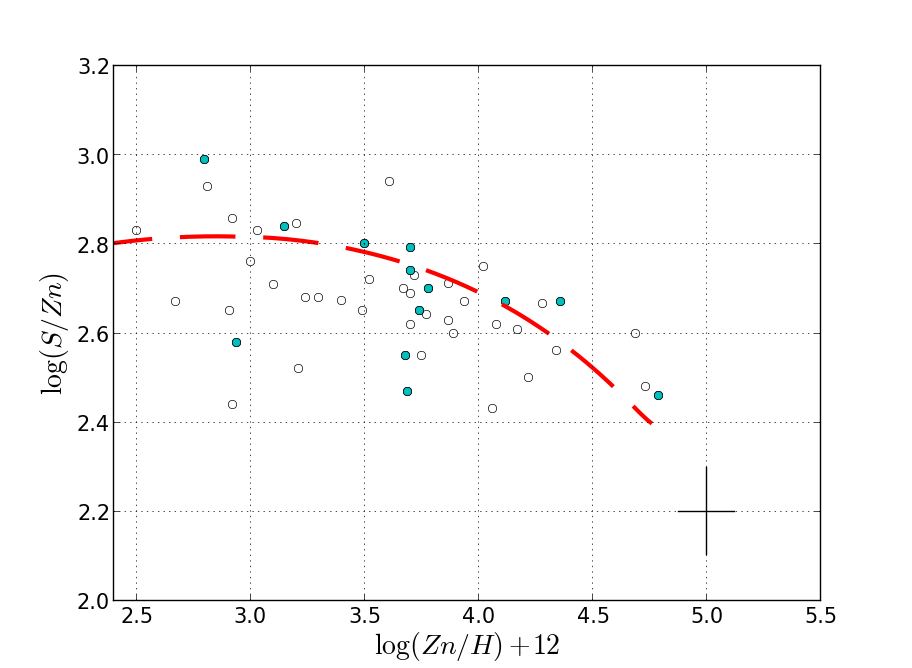}
\caption{Volatile S/Zn abundance ratios versus Zn/H in DLA systems. The long dashed red line represents the chemical pattern calculated for model I3, whose parameters are reported in Table~\ref{dust_param}. Data: open circles are the measurements which were already present in Vladilo et al. (2011) work, while cyan circles represent the values of the sample presented in Table~\ref{tab:DLA_sample}. In the figure, zinc and sulfur column densities are corrected as suggested by Kisielius et al. (2014; 2015) (see section~\ref{sec_DLA_data}).}
\label{fig:S_Zn}
\end{figure}

As a first step of our procedure, we tailored our model to match the observed S/Zn ratio, which is not affected by dust parameters. 
We performed some tests on the input parameters of chemical evolution models: we changed the wind parameter, star formation efficiency, infall mass and IMF, as already explained in section~\ref{dust_model_comparison}.
In  Fig.~\ref{fig:S_Zn} we show the comparison between the model results of the dwarf irregular and the observed abundances of volatile elements in DLA systems. The parameters for the model are reported in Table~\ref{dust_param}. 
We refer to V11 for a study of the impact of parameter variations
on the spread of the predicted S/Zn abundance ratios. 
To reproduce the full span of S/Zn values of DLAs as a function of metallicity,
the use of an inhomogeneous chemical evolution model would be required, similar to the one presented in Cescutti et al. (2008), and
such a task is beyond the aim of the present paper.

\begin{figure*}
\centering
\includegraphics[width=1.0\textwidth]{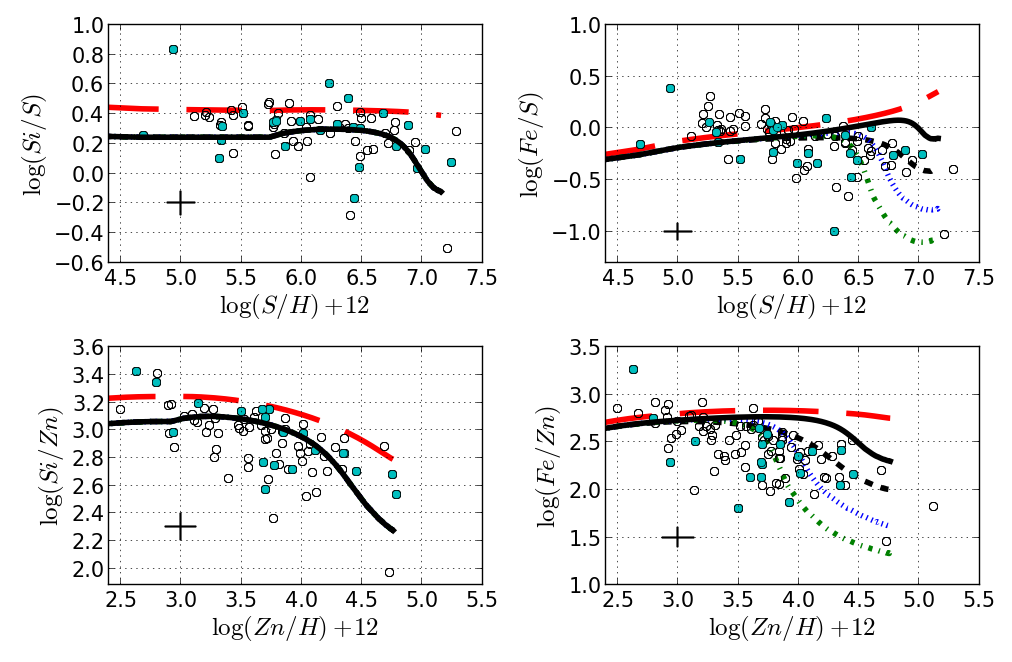}
\caption{Abundance ratio of refractory elements Si and Fe over volatile elements S (top panels) and Zn (bottom panels), against the absolute abundances of the volatile element. 
The red long-dashed lines represent the cosmic ISM abundance ratios of model I3, whereas the black solid lines represent the same model after the removal of the dust, i.e. the gas model. 
The black short-dashed lines represent a model including iron dust production from Type Ia SNe.
Blue dotted and green dashed-dotted lines show the predictions computed assuming a 5 and 10 times more efficient iron accretion, respectively. Data: same as Figure~\ref{fig:S_Zn}}
\label{fig:ref_vol}
\end{figure*}

In Fig.~\ref{fig:ref_vol} we show the relative abundance ratios between refractory (Si, Fe) and volatile (S,Zn) versus the absolute abundance of the corresponding volatile. We studied each possible combination of these elements: Si/S and Fe/S versus S (in the top panels), and Si/Zn and Fe/Zn versus Zn (bottom panels). For dust prescriptions we adopted more recent P11 dust condensation efficiencies, swept up mass for destruction as in Eq. (\ref{sweptup}) and the accretion as in Eq. (\ref{Asano_accr}). For a direct comparison with data, we remove the dust contribution from the chemical predictions of the ISM (red line), obtaining the \textit{gas model} (black solid lines). 

%%%%%%%%%%%%%%%%%%%%%%%%%%%%%%%
%%%%%%%%%%%%%%%%%%%%%%%%%%%%%%%
%%%%%%%%%%%%%%%%%%%%%%%%%%%%%%%
%%%%%%%%%%%%%%%%%%%%%%%%%%%%%%%
Assuming that volatile elements totally stay in the gas phase while a certain fraction of refractories is incorporated in dust grains, refractory gas abundances should show smaller values with respect to the total ISM (gas plus dust). For this reason we expect the refractory to volatile abundance ratios in the gas to lie below the total ISM abundances. 
In Fig.~\ref{fig:ref_vol} we see that the ISM model lies above the majority of the data, in agreement with this expectation. The cases where the measurements lie above the model could be due either to the natural dispersion of the DLA sample or to the uncertainties related to the stellar yields for zinc, sulfur and silicon, which are especially critical at low metallicities (Romano et al. 2010).
We obtain interesting and different results for silicon and iron.

%%%%%%%%%%%%%%%%%%%%%%%%%%%%%%%
%%%%%%%%%%%%%%%%%%%%%%%%%%%%%%%
%%%%%%%%%%%%%%%%%%%%%%%%%%%%%%%
%%%%%%%%%%%%%%%%%%%%%%%%%%%%%%%

In the left panels of Fig.~\ref{fig:ref_vol} we show the case of silicon. 
The measured gas-phase DLA data show a moderate decrease with increasing metallicity. This suggests that the amount of silicon in dust has a moderate tendency to increase in the course of galactic chemical evolution.  
Concerning the models, the difference of the predictions before and after the removal of the dust is evident (red and black solid lines, respectively).  We obtain a reasonable agreement between the model and the data for both Si/Zn and Si/S ratios. This suggests that the dust contribution of accretion together with dust production by Type II SNe and AGB is able to explain the depletion pattern of silicon observed in DLA systems. 
We notice that silicon observations are depleted even at the lowest metallicities of the sample, in agreement with a scenario in which Type II SNe give a fast contribution to silicon dust production.

In the right panels of the same Figure we show the pattern of iron abundance versus volatile elements. 
The abundance ratios show a marked decrease with metallicity as reported in previous work (Vladilo et al. 2011). 
In this case the observed iron depletion tends to vanish at the lowest metallicities, suggesting that the mechanisms of production of iron-rich dust take place on longer time scales than those typical of Type II SNe.
The gap between the total ISM model and the gas-phase data increases with metallicity, indicating that the mechanisms of production of iron-rich dust must be metallicity-dependent. 
However, at variance with the case of silicon, the gas-phase model (black solid line) does not fit at all the gas-phase Fe/S and Fe/Zn data. In fact, the model lies much higher than the data, suggesting that the adopted model predicts too little iron in dust. 
In this model, only the accretion process plays a significant contribution, whereas the dust production by Type II SNe and AGB stars leads to a negligible fraction with respect to the total iron abundance in the ISM. Even if we invoke a major contribution by either Type II SNe or AGB stars, the total iron in dust would be still negligible with respect to the huge iron amount ejected by Type I a SNe in the gas phase. For this reason, the model cannot predict any appreciable iron depletion until the metallicity becomes high enough to make the accretion process important.

It is evident that an extra source of iron dust production must be added to the model
to reduce the discrepancy with the data. We considered two possibilities.
First, we tested the potential contribution of Type Ia SNe  to iron dust using the D98 prescription, even if we know that there is no observational support for this hypothesis (see Section~\ref{sformation}). 
The results are shown as black short dashed lines in  Fig.~\ref{fig:ref_vol}.
One can see that this model is unable to follow the trend of the measured iron abundance ratios. 
As a second possibility, we assumed that
the bulk of iron is incorporated in a solid component, different from silicates, 
characterized by a high accretion efficiency. 
The existence of an iron dust population is suggested by other work (see session~\ref{discussion}). In particular, the existence of iron-rich, metallic nano-particles is  considered in recent studies (Draine $\&$ Hensley, 2012-2013). 
Such particles might have sizes one order of magnitude smaller than the standard size
of silicate particles adopted in Eq.~(\ref{Asano_accr}). The same equation predicts 
that the accretion time scale should be correspondingly smaller.  
Therefore, to increase the efficiency of iron accretion we reduced the accretion time-scale
by a factor 5 and 10. With such prescriptions we obtain a better match to the data 
(green short dotted lines and blue dashed-dotted lines in Fig.~\ref{fig:ref_vol}, respectively).

All the models that we have considered trace each other in the low metallicity range (early epochs), while they evolve in a substantially different way at metallicities above $\log (S/H)+12\simeq6.5$ and $\log (Zn/H)+12\simeq3.5$ for sulfur and zinc respectively. 
This is consistent with Eq. (\ref{Asano_accr}) and what we explained in section~\ref{following_paragraph}: the dust accretion becomes important as the metallicity increases and, in particular, when it reaches a critical value.

\subsection{Discussion}\label{discussion}

In our work we have reproduced the depletion pattern of silicon and iron in the ISM of DLA systems: we show that when we consider dust accretion and formation from Type II SNe and AGB stars, a good agreement is obtained for silicon, but not for iron.
The results that we have found support a scenario in which iron and silicon undergo a different history of dust formation and evolution.
Iron and silicon are believed to be coupled in silicate species, mostly in olivine ($Mg_{2y}Fe_{2(1-y)}SiO_4$) and pyroxene ($ Mg_{x}Fe_{(1-x)}SiO_3$) compounds.
The absorption of silicate features at 9.7 and 20 $\mu$m has been observed in a variety of environments, such as the diffuse ISM, cold and dark interstellar clouds, planetary nebulae and also in DLA systems (Nuth $\&$ Hecht 1990, Swamy 2005, Kulkarni et al.2007). 
Observations in the local interstellar clouds by Kimura et al. (2003) suggested an iron enrichment in the cores of silicate grains composed by triolite (FeS), kamacite (FeNi) or oxides (FeO). 
Additional interstellar observations suggest the existence of a dust species decoupled from silicates: Fe and Si depletion in the Small Magellanic Cloud often diverges (Sofia et al. 2006), indicating that iron is not tied to the same grains as silicon. Voshchinnikov et al. (2010) studied dust depletion in 196 different sight lines of the Milky Way, arguing that silicates grains cannot be a composition of olivines and pyroxenes only, but some amount of iron should reside in another dust population. 
Iron needles could represent an important additional dust species, having an appreciable contribution in the total amount of iron dust: Dwek (2004) argued that iron needles contribute to the unexpected extinction law in the mid-infrared observation ($3-8\mu m$) of the Galactic center. 
In addition, theoretical prescriptions (Hoyle 1999), different abundance ratio in various physical conditions (Voshchinnikov et al. 2010) and possible contribution of iron needles in Cas A (Dwek 2004) and SN 1987A (Wickramasinghe $\&$ Wickramasinghe 1993) may suggest that: 1) those needles can be readily created in SNe environments and, 2) the destruction of silicates grains in the warm medium is more effective than for Fe-rich grains or, in other words that iron particles are more resistant in the harshest ISM conditions. 

Whilst the existence of a form of iron dust decoupled from silicates
is suggested by many authors, its origin and nature are still under debate (Vladilo 2004). 
The iron dust problem arises from the fact the bulk of this element is produced by Type Ia SNe, but at the same time there is no evidence of iron dust particles in these SNe (see end of section~\ref{sformation}).  
In our work we suggest that this iron species may originate directly in the ISM. 
Further evidence of iron dust accreting in the ISM is provided by Dwek (2016). 
Here, we demonstrate this possibility in a chemical evolution context.
Draine $\&$ Hensley (2012-2013) also investigated the same possibility: 
they show that the sub millimeter and millimeter excess observed in low-metallicity galaxies might be explained by the presence of magnetic nano-particles, with radius $a<0.1\mu m$, which include a large fraction of interstellar Fe.

If future observations will prove the existence of metallic nano-particles, the possibility
that such solid component is partly produced by Type Ia SNe 
could be addressed with  specific observational tests. As far as models are concerned,
new algorithms for the production of metallic nano-particles by Type Ia SNe may be added 
to the efficient accretion in the ISM in order to improve the fit to the 
observed gas phase abundances in DLA systems. This possibility will be explored in a future work.

\section{Conclusions}\label{sec_conclusions}
In this work we have presented a chemical evolution model which takes into account the presence of dust utilizing new updated prescriptions. Dust formation is treated in the same way as first done by Dwek (1998), but with the inclusion of improved condensation efficiencies of Piovan et al. (2011). With respect to other models such as those of Calura et al. (2008) or Grieco et al. (2014), we have also changed the accretion and destruction prescriptions, which are two very important processes in dust evolution. We have applied our model to dwarf irregular galaxies and DLA systems. 
Our main results can be summarized as follows:
\begin{enumerate}
 \item We studied the dust production rate and the processes occurring in the ISM during the galactic lifetime of a typical irregular galaxy. We have computed the evolution of dust by considering dust production (Type II SNe, AGB stars), destruction and accretion processes. It is worth noting that we excluded the Type Ia SNe as dust producers since there is no observational evidence for that. We have found that dust accretion plays a fundamental role in dust evolution and in certain phases it becomes the dominant process, governing the evolution of the dust mass in the ISM, as predicted by Asano et al. (2013). 
 Moreover our model reproduces the observed dust-to-gas ratios as function of metallicity in such galaxies. 
 \item We investigated the impact of the cut-off of high mass stars (from $18$ to $80Mo$) on the chemical evolution of a typical irregular galaxy.
 We fail in reproducing the metallicity values observed in dwarf irregulars when the cut-off mass is assumed to be in the range 18-25$M_{\odot}$.
 On the other hand, this effect does not deeply affect the predicted range of dust-to-gas ratio.
 \item We compared the dust formation when both P11 and D98 condensation efficiencies are adopted. We found that the rate production of carbon is almost the same using different prescriptions, while the main differences concern silicates: using D98 condensation efficiencies, Type Ia SNe play a significant role and, in addition, a major contribution is given by Type II and LIMS. 
 \item Dust destruction represents a negligible process in dwarf irregulars, whereas the galactic wind is an important mechanism which can affect dust evolution: we showed that it can be the main responsible for stopping the accretion process in the ISM.
 \item We compared our model for irregulars with the data of DLA systems and we found that these objects can indeed be irregular galaxies, as already suggested in previous papers. We found a particular combination of parameters which best fit the DLAs.
 In particular, our comparison shows that the depletion pattern of silicon in these objects is well reproduced by the dust contributions of Type II SNe, AGBs and by the accretion process. 
 \item In the case of iron, at variance with the case of silicon, we find a good agreement with the data only when an extra dust source is considered: in particular, we tested the case of dust production by Type Ia SNe and the case of a more efficient accretion in the ISM. 
 The different behavior of iron and silicon that we find brings new evidence that a significant fraction of iron has to be incorporated into a dust population different from silicates, as suggested by previous works.
 Furthermore, as part of iron dust should be decoupled from silicates, it is possible that such species could originate in a different way: in particular, our results are consistent with a metallicity-dependent accretion of iron nano-particles. 
 \end{enumerate}

\section*{Acknowledgements}
We wish to thank I. J. Danziger for the useful discussion about dust in SN remnants, and also E. Spitoni F. Vincenzo and E. Gjergo for many fruitful discussions. 
FM and FC acknowledge financial support from PRIN-MIUR 2010-2011 project, "The Chemical and dynamical Evolution of the MW and Local Group Galaxies", prot. 2010LY5N2T.
We also thank an anonymous referee for valuable suggestions which improved the paper.

\begin{table*}
\begin{center}
\begin{tabular}{l c c c c c c c c c}
\hline
QSO & zab & $\log \left(\dfrac{N(Zn_{II})}{cm^{-2}}\right)$ & $\log \left(\dfrac{N(S_{II})}{cm^{-2}}\right)$ & $\log \left(\dfrac{N(Fe_{II})}{cm^{-2}}\right)$ &  $\log \left(\dfrac{N(Si_{II})}{cm^{-2}}\right)$ &Zn ref&S ref&Si ref&Fe ref\\
 \hline 
0008-0958	 &  1.7675  &  13.31 $\pm$ 0.05	 &  15.84 $\pm$ 0.05  &  15.62 $\pm$ 0.05	 &  16.04 $\pm$ 0.05	 &  35B	 &  35B	 &  35B	 &  35B	  \\
0142-100	 &  1.6265  &  11.43 $\pm$ 0.15	 &  14.53 $\pm$ 0.10  &  14.59 $\pm$ 0.03	 &  14.75 $\pm$ 0.03	 &  30a	 &  30a	 &  30a	 &  30a	  \\
0927+1543	 &  1.7311  &  13.38 $\pm$ 0.05	 &  --	 	      &  15.14 $\pm$ 0.24	 &  15.99 $\pm$ 0.05	 &  35B	 &  -	 &  35B	 &  35B	  \\
0927+5823	 &  1.6352  &  13.29 $\pm$ 0.05	 &  15.61 $\pm$ 0.05  &  --	 		 &  15.72 $\pm$ 0.05	 &  35B	 &  35B	 &  -	 &  35B	  \\
1013+5615	 &  2.2831  &  13.56 $\pm$ 0.05	 &  --	 	      &  --	 		 &  16.14 $\pm$ 0.05	 &  35B	 &  -	 &  -	 &  35B	  \\
1049-0110	 &  1.6577  &  13.14 $\pm$ 0.05	 &  15.47 $\pm$ 0.05  &  15.17 $\pm$ 0.05	 &  15.80 $\pm$ 0.05	 &  35B	 &  35B	 &  35B	 &  35B	  \\
1111-152	 &  3.266   &  12.32 $\pm$ 0.10	 &  14.62 $\pm$ 0.04  &  14.65 $\pm$ 0.03	 &  15.10 $\pm$ 0.07	 &  26e	 &  34Z	 &  34Z	 &  34Z	  \\
1155+0530	 &  3.326   &  12.89 $\pm$ 0.07	 &  15.40 $\pm$ 0.05  &  15.37 $\pm$ 0.05	 &  15.94 $\pm$ 0.05	 &  35B	 &  35B	 &  35B	 &  35B	  \\
1240+1455	 &  3.1078  &  12.90 $\pm$ 0.07	 &  15.56 $\pm$ 0.02  &  14.60 $\pm$ 0.03	 &  15.93 $\pm$ 0.03	 &  30d	 &  30d	 &  30d	 &  30d	  \\
1310+5424	 &  1.8006  &  13.57 $\pm$ 0.05	 &  --	 	      &  15.64 $\pm$ 0.05	 &  16.44 $\pm$ 0.05	 &  35B	 &  -	 &  35B	 &  35B	  \\
1337+3152	 &  3.1745  &  12.26 $\pm$ 0.26	 &  15.11 $\pm$ 0.2   &  14.91 $\pm$ 0.08	 &  15.50 $\pm$ 0.15	 &  30c	 &  30c	 &  30c	 &  30c	  \\
1454+0941	 &  1.7884  &  12.72 $\pm$ 0.05	 &  15.25 $\pm$ 0.06  &  15.02 $\pm$ 0.12	 &  15.47 $\pm$ 0.05	 &  35B	 &  35B	 &  35B	 &  35B	  \\
1552+4910	 &  1.9599  &  12.93 $\pm$ 0.05	 &  15.34 $\pm$ 0.05  &  15.47 $\pm$ 0.05	 &  15.98 $\pm$ 0.05	 &  35B	 &  35B	 &  35B	 &  35B	  \\
1604+3951	 &  3.1633  &  13.00 $\pm$ 0.10	 &  15.70 $\pm$ 0.02  &  15.40 $\pm$ 0.15	 &  16.09 $\pm$ 0.02	 &  30d	 &  30d	 &  30d	 &  30d	  \\
1610+4724	 &  2.5066  &  13.56 $\pm$ 0.05	 &  --	 	      &  15.62 $\pm$ 0.05	 &  16.16 $\pm$ 0.05	 &  35B	 &  -	 &  35B	 &  35B	  \\
1629+0913	 &  1.9023  &  12.68 $\pm$ 0.08	 &  15.24 $\pm$ 0.05  &  --	 		 &  15.32 $\pm$ 0.06	 &  35B	 &  35B	 &  -	 &  35B	  \\
1755+578	 &  1.9692  &  13.85 $\pm$ 0.05	 &  --	 	      &  15.79 $\pm$ 0.05	 &  16.58 $\pm$ 0.05	 &  35B	 &  -	 &  35B	 &  35B	  \\
1759+7539	 &  2.625   &  12.56 $\pm$ 0.10	 &  15.21 $\pm$ 0.02  &  14.94 $\pm$ 0.02	 &  15.55 $\pm$ 0.06	 &  22k	 &  19c	 &  19c	 &  22k	  \\
2132-4321	 &  1.916   &  12.69 $\pm$ 0.02	 &  --	 	      &  15.06 $\pm$ 0.04	 &  15.57 $\pm$ 0.02	 &  35A	 &  -	 &  35A	 &  35A	  \\
1142+0701	 &  1.8407  &  13.29 $\pm$ 0.05	 &  --	 	      &  15.47 $\pm$ 0.05	 &  --	 		 &  35B	 &  -	 &  35B	 &  -	  \\
1313+1441	 &  1.7947  &  13.30 $\pm$ 0.05	 &  --	 	      &  15.55 $\pm$ 0.05	 &  --	 		 &  35B	 &  -	 &  35B	 &  -	  \\
1417+4132	 &  1.9509  &  13.55 $\pm$ 0.05	 &  --	 	      &  15.58 $\pm$ 0.05	 &  --	 		 &  35B	 &  -	 &  35B	 &  -	  \\
0027-1836	 &  2.402   &  12.79 $\pm$ 0.02	 &  15.23 $\pm$ 0.02  &  14.97 $\pm$ 0.04	 &  15.67 $\pm$0.08	 &  27f	 &  27f	 &  28b	 &  28b	  \\
0642-5038	 &  2.659   &  12.75 $\pm$ 0.05	 &  15.35 $\pm$ 0.05  &  14.91 $\pm$ 0.03	 &  15.22 $\pm$0.06	 &  34a	 &  34a	 &  34Z	 &  34Z	  \\
1209+0919	 &  2.5841  &  12.98 $\pm$ 0.05	 &  --  	 	 &  15.25 $\pm$ 0.03	 &  15.91 $\pm$0.02	 &  27h	 &  -	 &  27h	 &  27h	  \\
0035-0918	 &  2.3401  &  --	 	 	 &  13.08 $\pm$ 0.1   &  12.96 $\pm$ 0.05	 &  13.37 $\pm$0.05	 &  -	 &  34B	 &  34B	 &  34B	  \\
0044+0018	 &  1.725   &  --	 	 	 &  15.27 $\pm$ 0.05  &  --	 		 &  15.34 $\pm$0.05	 &  -	 &  35B	 &  -	 &  35B	  \\
0142+0023	 &  3.3477  &  --	 	 	 &  13.28 $\pm$ 0.06  &  13.70 $\pm$ 0.10	 &  14.15 $\pm$0.03	 &  -	 &  30d	 &  30d	 &  30d	  \\
0234-0751	 &  2.3182  &  --	 	 	 &  14.18 $\pm$ 0.03  &  14.18 $\pm$ 0.03	 &  14.32 $\pm$0.09	 &  -	 &  34A	 &  34A	 &  34A	  \\
0450-13	 	 &  2.067   &  --	 	 	 &  14.28 $\pm$ 0.12  &  14.30 $\pm$ 0.07	 &  14.68 $\pm$0.10	 &  -	 &  26a	 &  26a	 &  26a	  \\
0958+0145	 &  1.9275  &  --	 	 	 &  14.44 $\pm$ 0.05  &  14.23 $\pm$ 0.05	 &  14.84 $\pm$0.06	 &  -	 &  35B	 &  35B	 &  35B	  \\
1024+0600	 &  1.895   &  --	 	 	 &  15.45 $\pm$ 0.05  &  15.27 $\pm$ 0.08	 &  15.81 $\pm$0.05	 &  -	 &  35B	 &  35B	 &  35B	  \\
1112+1333	 &  2.2709  &  --	 	 	 &  13.69 $\pm$ 0.09  &  13.59 $\pm$ 0.02	 &  13.95 $\pm$0.02	 &  -	 &  34B	 &  34B	 &  34B	  \\
1211+0422	 &  2.3766  &  --	 	 	 &  14.53 $\pm$ 0.04  &  14.62 $\pm$ 0.03	 &  14.91 $\pm$0.04	 &  -	 &  28k	 &  28k	 &  28k	  \\
1335+0824	 &  1.856   &  --	 	 	 &  15.29 $\pm$ 0.05  &  --	 		 &  15.73 $\pm$0.05	 &  -	 &  35B	 &  -	 &  35B	  \\
1340+1106	 &  2.7958  &  --	 	 	 &  14.22 $\pm$ 0.02  &  14.32 $\pm$ 0.02	 &  14.58 $\pm$0.02	 &  -	 &  31a	 &  31a	 &  31a	  \\
1509+1113	 &  2.0283  &  --	 	 	 &  15.69 $\pm$ 0.05  &  15.48 $\pm$ 0.07	 &  16.04 $\pm$0.05	 &  -	 &  35B	 &  35B	 &  35B	  \\
1004+0018	 &  2.5397  &  --	 	 	 &  15.09 $\pm$ 0.02  &  15.13 $\pm$ 0.02	 &  --	 		 &  -	 &  34A	 &  34A	 &  -	  \\
1009+0713	 &  0.114   &  --	 	 	 &  15.25 $\pm$ 0.12  &  15.29 $\pm$ 0.17	 &  --	 		 &  -	 &  31b	 &  31b	 &  -	  \\
1451+1223	 &  2.255   &  11.85 $\pm$ 0.11	 &  --	 	      &  14.33 $\pm$ 0.07	 &  --	 		 &  23b	 &  -	 &  -	 &  23b	  \\
0112+029	 &  2.423   &  --	 	 	 &  14.83 $\pm$ 0.08  &  14.86 $\pm$ 0.05	 &  --	 		 &  -	 &  23d	 &  -	 &  23d	  \\
1036-2257	 &  2.777   &  --	 	 	 &  14.79 $\pm$ 0.02  &  14.68 $\pm$ 0.02	 &  --	 		 &  -	 &  23g	 &  -	 &  23g	  \\
0000-263  	 &  3.3901  &  12.01 $\pm$ 0.05 	 &  14.70 $\pm$ 0.03  &  14.76 $\pm$ 0.03  	 &  15.06 $\pm$ 0.02   	 & 20c   &  16d   &  20c   &  20c        \\  
0010-0012 	 &  2.0250  &  12.25 $\pm$ 0.05  	 &  14.96 $\pm$ 0.05  &	 15.06 $\pm$ 0.05    	 &  15.31 $\pm$ 0.05   	 & 23d   &  25d   &  25d   &  23d   	 \\
0058-2914 	 &  2.6711  &  12.23 $\pm$ 0.05  	 &  14.92 $\pm$ 0.03  &  14.75 $\pm$ 0.05  	 &  15.23 $\pm$ 0.07   	 & 23d   &  25d   &  25d   &  23d  	 \\
0100+130  	 &  2.3090  &  12.47 $\pm$ 0.10  	 &  15.09 $\pm$ 0.06  &	 13.37 $\pm$ 0.01  	 &    -             	 & 24a   &  24a   &  -     &  24a	 \\
0102-1902 	 &  2.3693  &  11.77 $\pm$ 0.11  	 &  14.30 $\pm$ 0.04  &  14.47 $\pm$ 0.10  	 &                  	 & 23d   &  25d   &  -     &  23d	 \\
0201+365  	 &  2.4620  &  12.76 $\pm$ 0.30  	 &  15.29 $\pm$ 0.02  &	 15.01 $\pm$ 0.01   	 &  15.53 $\pm$ 0.01   	 & 16e   &  22j   &  22j   &  22j      	 \\
0216+080  	 &  2.2931  &  12.47 $\pm$ 0.05  	 &  15.04 $\pm$ 0.02  &	 14.88 $\pm$ 0.02  	 &  15.45 $\pm$ 0.04   	 & 26e   &  31P   &  16d   &  31P	 \\
0347-383  	 &  3.0250  &  12.23 $\pm$ 0.12  	 &  14.76 $\pm$ 0.05  &  14.43 $\pm$ 0.01   	 &  14.77 $\pm$ 0.04   	 & 23d   &  25d   &  23d   &  22e      	 \\
0405-443  	 &  2.5505  &  12.44 $\pm$ 0.05  	 &  14.82 $\pm$ 0.06  &  14.95 $\pm$ 0.06   	 &  15.32 $\pm$ 0.04   	 & 23e   &  23e   &  23e   &  23e   	 \\
0528-2505 	 &  2.1410  &  12.29 $\pm$ 0.03  	 &  14.83 $\pm$ 0.04  &  14.85 $\pm$ 0.09  	 &  15.22 $\pm$ 0.05   	 & 26e   & 23a    &  23a   &  23a        \\
0812+32   	 &  2.6260  &  13.15 $\pm$ 0.02  	 &  15.63 $\pm$ 0.08  &  15.98 $\pm$ 0.05  	 &  15.09 $\pm$ 0.01   	 & 27h   & 27h    &  27h   &  27h        \\
0841+129  	 &  2.3745  &  12.20 $\pm$ 0.05  	 &  14.77 $\pm$ 0.03  &  14.87 $\pm$ 0.04  	 &  15.21 $\pm$ 0.04   	 & 23a   & 23a    &  27b   &  23a	 \\
0841+129  	 &  2.4764  &  11.69 $\pm$ 0.10  	 &  14.48 $\pm$ 0.10  &  14.50 $\pm$ 0.05  	 &  14.99 $\pm$ 0.03   	 & 26a   & 26a    &  26a   &  26a	 \\
0953+5230 	 &  1.7680  &  12.89 $\pm$ 0.05  	 &  15.35 $\pm$ 0.05  &  14.99 $\pm$ 0.05  	 &  15.67 $\pm$ 0.05   	 & 26k   & 26k    &  26k   &  26k	 \\
 \hline
\end{tabular}
\caption{The DLA sample used in this work. In the first column the QSO name, in the second the absorption redshift of the DLA system. Column density measurements of ZnII, SII, FeII and SiII are reported in column 3,4,5 and 6, respectively. 
Reference codes for the column densities are reported in columns 7,8,9 and 10: we report the corresponding literature references in Table~\ref{tab:references}} 
\label{tab:DLA_sample}
\end{center}
\end{table*}

\begin{table*}
\begin{center}
\begin{tabular}{l c c c c c c c c c}
\hline
QSO & zab & $\log \left(\dfrac{N(Zn_{II})}{cm^{-2}}\right)$ & $\log \left(\dfrac{N(S_{II})}{cm^{-2}}\right)$ & $\log \left(\dfrac{N(Fe_{II})}{cm^{-2}}\right)$ &  $\log \left(\dfrac{N(Si_{II})}{cm^{-2}}\right)$ &Zn ref&S ref&Si ref&Fe ref\\
 \hline 
1116+4118 	 &  2.9422  &  12.40 $\pm$ 0.33  	&  15.01 $\pm$ 0.10  &  14.69 $\pm$ 0.04   	 &  15.34 $\pm$ 0.08   	 & 27j   & 27j    &  27j   &  27j	 \\
1210+1731 	 &  1.8918  &  12.40 $\pm$ 0.05  	&  14.96 $\pm$ 0.03  &  15.01 $\pm$ 0.03   	 &  15.33 $\pm$ 0.03   	 & 27b   & 27b    &  27b   &  27b  	 \\
1223+178  	 &  2.4661  &  12.42 $\pm$ 0.05  	&  15.14 $\pm$ 0.04  &  15.21 $\pm$ 0.05   	 &  15.50 $\pm$ 0.03  	 & 25d   & 23d    &  25d   &  23d	 \\
1331+170  	 &  1.7764  &  12.54 $\pm$ 0.02  	&  15.08 $\pm$ 0.11  &  14.63 $\pm$ 0.03   	 &  15.30 $\pm$ 0.01   	 & 24a   & 24a    &  24a   &  24a	 \\
2138-4427 	 &  2.8510  &  11.99 $\pm$ 0.05  	&  14.50 $\pm$ 0.02  &  14.65 $\pm$ 0.05  	 &  14.86 $\pm$ 0.02   	 & 23d   & 25d    &  25d   &  23d	 \\
2206-199  	 &  1.9200  &  12.95 $\pm$ 0.02  	&  15.42 $\pm$ 0.02  &  15.31 $\pm$ 0.01    	 &  15.80 $\pm$ 0.01  	 & 21i   & 31P    &  ww1   &  31P	 \\
2222-0946 	 &  2.3540  &  12.83 $\pm$ 0.01  	&  15.31 $\pm$ 0.01  &  15.13 $\pm$ 0.01    	 &  15.62 $\pm$ 0.01     & 33a   & 33a    &  33a   &  33a	 \\
2230+025  	 &  1.8642  &  12.80 $\pm$ 0.11  	&  15.29 $\pm$ 0.10  &  15.25 $\pm$ 0.05   	 &  15.70 $\pm$ 0.05     & 26a   &  26a   &  26a   &  26a	 \\
2231-0015 	 &  2.0662  &  12.30 $\pm$ 0.05  	&  15.10 $\pm$ 0.15  &  14.83 $\pm$ 0.03    	 &  15.29 $\pm$ 0.04     & 24a   & 24a    &  24a   &  24a	 \\
2243-6031 	 &  2.3300  &  12.47 $\pm$ 0.02  	&  15.02 $\pm$ 0.03  &  14.92 $\pm$ 0.03   	 &  15.36 $\pm$ 0.02     & 26f   & 22f    &  22f   &  22f	 \\
2314-409  	 &  1.8573  &  12.52 $\pm$ 0.10  	&  15.10 $\pm$ 0.15  &  15.08 $\pm$ 0.10  	 &  15.41 $\pm$ 0.10     & 21c   &  21c   &  21c   &  21c 	 \\
2318-1107 	 &  1.9890  &  12.50 $\pm$ 0.06  	&  15.09 $\pm$ 0.04  &  14.91 $\pm$ 0.04  	 &  15.34 $\pm$ 0.04     & 27f   &  27f   &  27f   & 27f 	 \\
2343+1232 	 &  2.4313  &  12.25 $\pm$ 0.10  	&  14.66 $\pm$ 0.05  &  14.52 $\pm$ 0.05   	 &  15.15 $\pm$ 0.06     & 27f   &  27f   &  27f   & 27f 	 \\
0013-004  	 &  1.9731  &  12.82 $\pm$ 0.04  	&  15.28 $\pm$ 0.02  &  14.84 $\pm$ 0.03   	 &  15.43 $\pm$ 0.03     & 22h   &  22h   &  22h   & 22h	 \\
0551-3637 	 &  1.9615  &  13.02 $\pm$ 0.05  	&  15.38 $\pm$ 0.11  &  15.05 $\pm$ 0.05        &  15.62 $\pm$ 0.06     & 22d   & 22d    &  22d   &  22d	 \\
0918+1636 	 &  2.5832  &  13.40 $\pm$ 0.02  	&  15.82 $\pm$ 0.02  &  15.43 $\pm$ 0.01   	 &  16.01 $\pm$ 0.01     & 31C   & 31C   &  31C   &  31C	 \\
1439+1117 	 &  2.4184  &  12.93 $\pm$ 0.04  	&  15.27 $\pm$ 0.06  &  14.28 $\pm$ 0.05  	 &  14.80 $\pm$ 0.04     & 28d   & 28d   &  28d   &  28d	 \\
1443+2724 	 &  4.2240  &  12.99 $\pm$ 0.03  	&  15.52 $\pm$ 0.02  &  15.33 $\pm$ 0.03  	 &  --              	 & 28c   & 26f   &  -     &  26f	 \\
1444+014  	 &  2.0870  &  12.12 $\pm$ 0.15  	&  14.62 $\pm$ 0.08  &  14.00 $\pm$ 0.06   	 &  14.38 $\pm$ 0.06   	 & 23d   & 23d   &  23d   &  23d	 \\
0149+33   	 &  2.1410  &  11.50 $\pm$ 0.10  	&   --    	      &  14.23 $\pm$ 0.02  	 &  14.57 $\pm$ 0.05   	 & ww1   & -     &  ww1   &  ww1	 \\
0203-0910 	 &  1.0280  &  13.15 $\pm$ 0.15  	&   --  	      &  15.65 $\pm$ 0.03   	 &  --              	 & 29d   & -     &  -     &  29d	 \\
0225+0054 	 &  2.7140  &  12.89 $\pm$ 0.10  	&   --    	      &  15.30 $\pm$ 0.10   	 &  15.62 $\pm$ 0.10   	 & 26c   & -     &  26c   &  26c	 \\
0256+0110 	 &  0.7250  &  13.19 $\pm$ 0.04  	&   --    	      &  15.13 $\pm$ 0.30  	 &  --         	    	 & 26j   & -     &  -     &  26j	 \\
0302-223  	 &  1.0095  &  12.45 $\pm$ 0.04  	&   --    	      &  14.67 $\pm$ 0.043   	 &  15.18 $\pm$ 0.04   	 & 20e   & -     &  20e   &  20e	 \\
0354-2724 	 &  1.4051  &  12.73 $\pm$ 0.03  	&   --    	      &  15.15 $\pm$ 0.05   	 &  --		    	 & 27e   & -     &  -     &  27e	 \\
0438-436  	 &  2.3474  &  12.72 $\pm$ 0.03  	&   --    	      &  14.95 $\pm$ NOERR   	 &  --              	 & 25a   & -     &  -     &  25a	 \\
0454+039  	 &  0.8597  &  12.42 $\pm$ 0.06  	&   --    	      &  15.17 $\pm$ 0.04    	 &  15.45 $\pm$ 0.09   	 & 28a   & -     &  20e   &  20e	 \\
0458-02   	 &  2.0400  &  13.13 $\pm$ 0.02  	&   --    	      &  15.40 $\pm$ 0.05     	 &  --		    	 & ww1   & -     &  -     &  19e	 \\
0515-4414 	 &  1.1510  &  12.11 $\pm$ 0.04  	&   --    	      &  14.24 $\pm$ 0.20    	 &  14.74 $\pm$ 0.18   	 & 20b   & -     &  20b   &  20b	 \\
0933+733  	 &  1.4790  &  12.71 $\pm$ 0.02  	&   --    	      &  15.19 $\pm$ 0.01    	 &  --          	 & 25e   & -     &  -     &  25e	 \\
0935+417  	 &  1.3726  &  12.25 $\pm$ 0.10  	&   --    	      &  14.82 $\pm$ 0.10 	 &  -- 		    	 & 15b   & -     &  -     &  15b	 \\
0948+433  	 &  1.2330  &  13.15 $\pm$ 0.02  	&   --    	      &  15.56 $\pm$ 0.01    	 &  --              	 & 25e   & -     &  -     &  25e	 \\
1010+0003 	 &  1.2651  &  13.01 $\pm$ 0.02  	&   --    	      &  15.26 $\pm$ 0.05   	 &  --   	    	 & 28a   & -     &  -     &  26g	 \\
1013+0035 	 &  3.1040  &  13.33 $\pm$ 0.02  	&   --    	      &  15.18 $\pm$ 0.050   	 &  15.78 $\pm$ 0.020  	 & 27h   & -     &  27h   &  27h	 \\
1055-301  	 &  1.9035  &  12.91 $\pm$ 0.03  	&   --    	      &  15.44 $\pm$ NOERR   	 &  15.95 $\pm$ 0.1NOER	 & 25a   & -     &  25a   &  25a	 \\
1104-1805 	 &  1.6616  &  12.48 $\pm$ 0.02  	&   --    	      &  14.77 $\pm$ 0.020 	 &  15.45 $\pm$ 0.020  	 & 19b   & -     &  19b   &  19b	 \\
1107+0048 	 &  0.7410  &  13.06 $\pm$ 0.15  	&   --    	      &  15.53 $\pm$ 0.02  	 &  --  	  	 & 26j   & -     &  -     &  26j	 \\
1116+4118	 &  2.6617  &  12.40 $\pm$ 0.20  	&   --    	      &  14.36 $\pm$ 0.10  	 &  15.05 $\pm$ 0.11   	 & 27j   & -     &  27j   &  27j	 \\
1117-1329 	 &  3.3511  &  12.25 $\pm$ 0.06  	&   --    	      &  14.82 $\pm$ 0.05   	 &  15.12 $\pm$ 0.04   	 & 22g   & -     &  22g   &  22g	 \\
1137+3907 	 &  0.7190  &  13.43 $\pm$ 0.05  	&   --    	      &  15.45 $\pm$ 0.05    	 &  --		    	 & 26g   & -     &  -     &  26g	 \\
1157+0128 	 &  1.9436  &  12.99 $\pm$ 0.05  	&   --    	      &  15.46 $\pm$ 0.02 	 &  15.97 $\pm$ 0.02   	 & 27b   & -     &  27b   &  27b	 \\
1215+33   	 &  1.9990  &  12.33 $\pm$ 0.05  	&   --    	      &  14.75 $\pm$ 0.05    	 &  15.03 $\pm$ 0.02   	 & ww1   & -     &  19e   &  ww1	 \\
1225+0035 	 &  0.7731  &  13.23 $\pm$ 0.07  	&   --     	      &  15.69 $\pm$ 0.03   	 &  --              	 & 28a   & -     &  -     &  26g	 \\
1230-101  	 &  1.9314  &  12.94 $\pm$ 0.05  	&   --     	      &  15.32 $\pm$ 0.10   	 &  15.77 $\pm$ 0.10     & 25a   & -     &  25a   &  25a	 \\
1249-0233 	 &  1.7810  &  13.11 $\pm$ 0.10  	&   --     	      &  15.47 $\pm$ 0.10   	 &  15.80 $\pm$ 0.10     & 26c   & -     &  26c   &  26c	 \\
1253-0228 	 &  2.7830  &  12.77 $\pm$ 0.07  	&   --     	      &  15.36 $\pm$ 0.04   	 &  --              	 & 23g   & -     &  -     &  23g	 \\
1323-0021 	 &  0.7160  &  13.43 $\pm$ 0.05  	&   --     	      &  15.15 $\pm$ 0.03   	 &  --              	 & 26i   & -     &  -     &  26i	 \\
1328+307  	 &  0.6922  &  12.53 $\pm$ 0.03  	&   --     	      &  15.09 $\pm$ 0.01   	 &  --              	 & 28j   & -     &  -     &  28j	 \\
1351+318  	 &  1.1491  &  12.52 $\pm$ 0.13  	&   --     	      &  14.74 $\pm$ 0.09   	 &  15.23 $\pm$ 0.13   	 & 19d   & -     &  19d   &  19d	 \\
1354+258  	 &  1.4200  &  12.59 $\pm$ 0.13  	&   --     	      &  15.03 $\pm$ 0.09   	 &  15.36 $\pm$ 0.13   	 & 19d   & -     &  19d   &  19d	 \\
1426+6039 	 &  2.8268  &  12.18 $\pm$ 0.04  	&   --     	      &  14.48 $\pm$ 0.01   	 &  --              	 & 27h   & -     &  -     &  27h	 \\
1501+0019 	 &  1.4832  &  12.93 $\pm$ 0.06  	&   --     	      &  --     		 &  15.71 $\pm$ 0.02   	 & 26g   & -     &  26g   &  -	 	 \\
1727+5302 	 &  0.9449  &  13.27 $\pm$ 0.05  	&   --     	      &  15.38 $\pm$ 0.140  	 &  15.94 $\pm$ 0.02   	 & 24c   & -     &  24c   &  24c	 \\
1727+5302 	 &  1.0311  &  12.65 $\pm$ 0.05  	&   --     	      &  14.54 $\pm$ 0.100  	 &  15.60 $\pm$ 0.03   	 & 24c   & -     &  24c   &  24c	 \\
1733+5533 	 &  0.9984  &  12.89 $\pm$ 0.06  	&   --     	      &  --                  	 &  15.48 $\pm$ 0.06   	 & 28a   & -     &  26g   &  -	 	 \\
1850+40   	 &  1.9900  &  13.35 $\pm$ 0.10  	&   --     	      &  15.58 $\pm$ 0.10   	 &  --              	 & 18g   & -     &  -     &  18g	 \\
2059-0528 	 &  2.2100  &  12.94 $\pm$ 0.10  	&   --     	      &  15.00 $\pm$ 0.10   	 &  15.36 $\pm$ 0.10   	 & 26c   & -     &  26c   &  26c	 \\
2228-3954 	 &  2.0950  &  12.51 $\pm$ 0.05  	&   --     	      &  15.17 $\pm$ 0.05   	 &  --              	 & 28c   & -     &  -     &  28c	 \\
2340-00   	 &  2.0545  &  12.63 $\pm$ 0.08  	&   --     	      &  --    	             	 &  15.17 $\pm$ 0.04   	 & 27h   & -     &  27h   &  -	 	 \\
2359-0216 	 &  2.0950  &  12.60 $\pm$ 0.03  	&   --     	      &  14.55 $\pm$ 0.03   	 &  15.40 $\pm$ 0.02   	 & 19e   & -     &  ww1   &  ww1	 \\
0135-273  	 &  2.8000  &  --    	         	&   14.78 $\pm$ 0.14 &  14.77 $\pm$ 0.11 	 &  --                	 & -     & 23d   &  -     &  23d	 \\
 \hline
\end{tabular}
\caption{Continues from Table~\ref{tab:DLA_sample}} 
\end{center}
\end{table*}

\begin{table*}
\begin{center}
\begin{tabular}{l c c c c c c c c c}
\hline
QSO & zab & $\log \left(\dfrac{N(Zn_{II})}{cm^{-2}}\right)$ & $\log \left(\dfrac{N(S_{II})}{cm^{-2}}\right)$ & $\log \left(\dfrac{N(Fe_{II})}{cm^{-2}}\right)$ &  $\log \left(\dfrac{N(Si_{II})}{cm^{-2}}\right)$ &Zn ref&S ref&Si ref&Fe ref\\
 \hline 
0201+1120 	 &  3.3848  &  --    	         	&   15.21 $\pm$ 0.11 &  15.35 $\pm$ 0.05   	 &  --              	 & -     & ww1   &  -     &  ww1	 \\
0242-2917 	 &  2.5600  &  --    	         	&   14.11 $\pm$ 0.04 &  14.36 $\pm$ 0.04   	 &  --              	 & -     & 28c   &  -     &  28c	 \\
0254-4025 	 &  2.0460  &  --    	         	&   14.10 $\pm$ 0.04 &  14.17 $\pm$ 0.04   	 &  --              	 & -     & 28c   &  -     &  28c	 \\
0255+00   	 &  3.9146  &  --    	         	&   14.72 $\pm$ 0.02 &  14.75 $\pm$ 0.15   	 &  --              	 & -     & ww1   &  -     &  ww1	 \\
0300-3152 	 &  2.1790  &  --    	         	&   14.20 $\pm$ 0.04 &  14.21 $\pm$ 0.04   	 &  --                	 & -     & 28c   &  -     &  28c	 \\
0336-0142 	 &  3.0621  &  --    	         	&   14.99 $\pm$ 0.02 &  14.91 $\pm$ 0.03   	 &  15.25 $\pm$ 0.03  	 & -     & 22j   &  26e   &  22j	 \\
0425-5214 	 &  2.2240  &  --    	         	&   14.07 $\pm$ 0.04 &  13.96 $\pm$ 0.04   	 &  --                	 & -     & 28c   &  -     &  28c	 \\
0741+4741 	 &  3.0174  &  --      		 	&   14.00 $\pm$ 0.02 &  14.05 $\pm$ 0.01   	 &  14.35 $\pm$ 0.01  	 & -     & ww1   &  22j   &  22j	 \\
0900+4215 	 &  3.2458  &  --      		 	&   14.65 $\pm$ 0.02 &  14.54 $\pm$ 0.02   	 &  --                	 & -     & 27h   &  -     &  27h	 \\
0957+33   	 &  4.1798  &  --    	         	&   14.39 $\pm$ 0.06 &  14.13 $\pm$ 0.05   	 &  14.56 $\pm$ 0.01  	 & -     & ww1   &  ww1   &  ww1	 \\
1021+3001 	 &  2.9490  &  --    	         	&   13.87 $\pm$ 0.07 &  14.04 $\pm$ 0.01   	 &  14.32 $\pm$ 0.02  	 & -     & 27h   &  27h   &  27h	 \\
1132+2243 	 &  2.7830  &  --    	         	&   14.07 $\pm$ 0.06 &  14.02 $\pm$ 0.02   	 &  14.49 $\pm$ 0.12  	 & -     & 23g   &  23g   &  23g	 \\
1220-1800 	 &  2.1130  &  --    	         	&   14.39 $\pm$ 0.03 &  14.35 $\pm$ 0.03   	 &  --                	 & -     & 28c   &  -     &  28c	 \\
1337+1121 	 &  2.7957  &  --    	         	&   14.33 $\pm$ 0.02 &  14.07 $\pm$ 0.02   	 &  14.79 $\pm$ 0.07  	 & -     & 27h   &  27h   &  27h	 \\
1354-1046 	 &  2.5010  &  --    	         	&   14.13 $\pm$ 0.10 &  14.35 $\pm$ 0.10   	 &  --                	 & -     & 28c   &  -     &  28c	 \\
1435+5359 	 &  2.3427  &  --    	         	&   14.78 $\pm$ 0.05 &  --       		 &  15.13 $\pm$ 0.02  	 & -     & 27c   &  27c   &  -	 	 \\
1558-0031 	 &  2.7026  &  --    	         	&   14.07 $\pm$ 0.02 &  --         		 &  14.24 $\pm$ 0.02  	 & -     & 27c   &  27c   &  -	 	 \\
2059-360  	 &  2.5073  &  --    	         	&   13.49 $\pm$ 0.23 &  13.53 $\pm$ 0.05   	 &  13.91 $\pm$ 0.03  	 & -     & 25d   &  25d   &  23d	 \\
2059-360  	 &  3.0830  &  --    	         	&   14.38 $\pm$ 0.05 &  14.52 $\pm$ 0.07   	 &  14.80 $\pm$ 0.05  	 & -     & 25d   &  25d   &  20d	 \\
2222-3939 	 &  2.1540  &  --    	         	&   14.08 $\pm$ 0.03 &  14.42 $\pm$ 0.03   	 &  --                	 & -     & 28c   &  -     &  28c	 \\
2241+1352 	 &  4.2820  &  --    	         	&   14.58 $\pm$ 0.03 &  14.76 $\pm$ 0.11   	 &  15.06 $\pm$ 0.05  	 & -     & 23g   &  23g   &  23g	 \\
2332-0924 	 &  3.0572  &  --    	         	&   14.13 $\pm$ 0.04 &  14.06 $\pm$ 0.03   	 &  14.64 $\pm$ 0.03  	 & -     & 34Z   &  34Z   &  34Z	 \\
2342+3417 	 &  2.9082  &  --    	         	&   15.19 $\pm$ 0.02 &  15.02 $\pm$ 0.06   	 &  15.62 $\pm$ 0.02  	 & -     & 27h   &  27h   &  27h	 \\
2348-147  	 &  2.2790  &  --    	         	&   13.75 $\pm$ 0.06 &  13.84 $\pm$ 0.05   	 &  14.18 $\pm$ 0.05  	 & -     & 26a   &  26a   &  26a	 \\
1232+0815 	 &  2.3377  &  --    	         	&   14.81 $\pm$ 0.09 &  14.44 $\pm$ 0.08   	 &  15.06 $\pm$ 0.05  	 & -     & 31c   &  31c   &  31c	 \\
2348-0108 	 &  2.4272  &  --    	         	&   15.06 $\pm$ 0.04 &  14.83 $\pm$ 0.03   	 &  15.26 $\pm$ 0.09  	 & -     & 27g   &  27g   &  27g    	 \\
 \hline
\end{tabular}
\caption{Continues from Table~\ref{tab:DLA_sample}} 
\end{center}
\end{table*}

\begin{table}
 \begin{center}
\begin{tabular}{l p{8cm} l p{8cm}}
\hline
\hline
Code & Reference & Code & Reference\\
\hline
16d & Lu, L., Sargent, W.L.W., Barlow, T.A., et al. 1996, ApJS, 107, 475 
& 26f & Ledoux, C., Petitjean, P., Srianand, R. 2006, ApJ, 640, L25 \\

16e & Prochaska, J.X., Wolfe, A.M. 1996, ApJ, 470, 403 
& 26g & Meiring, J.D., Kulkarni, V.P., Khare, P., et al. 2006, MNRAS, 370, 43 \\

18g &  Prochaska, J.X., Wolfe, A.M. 1998, ApJ, 507, 113 
& 26i & Peroux, C., Kulkarni, V.P., Meiring, J., et al. 2006, A$\&$A, 450, 53  \\

19b & Lopez, S., Reimers, D., Rauch, et al. 1999, ApJ, 513, 598
& 26j & Peroux, C., Meiring, J.D., Kulkarni, V.P., et al. 2006, MNRAS, 372, 369 \\

19c & Outram, P.J., Chaffee, F.H., Carswell, R.F. 1999, MNRAS, 310, 289 
& 26k & Prochaska, J.X., O'Meara, J.M., Herbert-Fort, S., et al. 2006, ApJ, 648, L97 \\

19e & Prochaska, J.X., Wolfe, A.M. 1999, ApJS, 121, 369 
& 27b & Dessauges-Zavadsky, M., Calura, F., Prochaska, et al. 2007, A$\&$A, 470, 431 \\

19d & Pettini, M., Ellison, S.L., et al. 1999, ApJ, 510, 576 
& 27e & Meiring, J.D., Lauroesch, J.T., Kulkarni, V.P., et al. 2007, MNRAS, 376, 557 \\

20b & de la Varga, A., Reimers, D., Tytler, D., et al. 2000, A$\&$A, 363, 69 
& 27f & Noterdaeme, P., Ledoux, C., Petitjean, P., et al. 2007, A$\&$A, 474, 393\\

20c & Molaro, P., Bonifacio, P., Centurion, M., et al. 2000, ApJ, 541, 54 
& 27g & Noterdaeme, P., Petitjean, P., Srianand, R., et al. 2007, A$\&$A, 469, 425 \\ 

20e & Pettini, M., Ellison, S., Steidel, C.C., et al. 2000, ApJ, 532, 65 
& 27h & Prochaska, J.X., Wolfe, A.M., Howk, J.C., et al. 2007, ApJS, 171, 29 \\

21i & Ellison, S.L., Ryan, S.G., Prochaska, J.X. 2001, MNRAS, 326, 628 &
27j & Ellison, S.L., Hennawi, J.F., et al. 2007, MNRAS, 378, 801 \\

22d &  Ledoux, C., Srianand, R., Petitjean, P. 2002, A$\&$A, 392, 781 
& 28a & Nestor,D.B., Pettini, M., et al. 2008, MNRAS, 390, 1670 \\

22e & Levshakov, S.A., Dessauges-Zavadsky, M., et al. 2002, ApJ, 565, 696 
& 28b & Noterdaeme, P., Ledoux, C., et al. 2008, A$\&$A, 481, 327 \\

22f & Lopez, S., Reimers, D., et al. 2002, A$\&$A, 385, 778 
& 28c & Noterdaeme, P., Ledoux, C., et al. 2008, A$\&$A, 481, 327 \\

22g & Peroux, C., Petitjean, P., et al. 2002, NewA, 7, 577 
& 28d & Noterdaeme, P., Petitjean, P., et al. 2008, A$\&$A, 491, 397 \\

22h & Petitjean, P., Srianand, R., Ledoux, C. 2002, MNRAS, 332, 383 
& 28j & Wolfe, A.M., Jorgenson, R.A., Robishaw, T., et al. 2008, Nature, 455, 638\\

22j & Prochaska, J.X., Henry, R.B.C., O'Meara, J.M., et al. 2002, PASP, 114, 933 
& 28k & Lehner, N., Howk, J.C., et al. 2008, MNRAS, 390, 2\\

22k & Prochaska, J.X., Howk, J.C., O'Meara, J.M., et al. N. 2002, ApJ, 571, 693 
& 29d & Monier, E.M., Turnshek, et al. 2009, AJ, 138, 1609 \\

23a & Centurion, M., Molaro, P., Vladilo, G., et al. 2003, A$\&$A, 403, 55 
& 30a & Cooke, R., Pettini, M., Steidel, C.C., et al. 2010, MNRAS, 409, 679 \\

23b & Dessauges-Zavadsky, M., Peroux, C., Kim, T.-S., et al. 2003, MNRAS, 35, 447 
& 30c & Srianand, R., Gupta, N., Petitjean, P., et al. 2010, MNRAS, 405, 1888\\

23d & Ledoux, C., Petitjean, P., Srianand, R. 2003, MNRAS, 346, 209 
& 30d & Ellison, S.L., Prochaska, J.X., Hennawi, J., et al. 2010, MNRAS, 406, 1435 \\

23e & Lopez, S., Ellison, S.L. 2003, A$\&$A, 403, 573 
& 31a & Cooke, R.,Pettini,M.,Steidel,C., et al. 2011, MNRAS,417,1534\\

23g & Prochaska, J.X., Gawiser, E., Wolfe, A.M., et al. 2003, ApJS, 147, 227 
& 31b & Meiring, J.D., Tripp, T.M., Prochaska, J.X., et al. 2011, ApJ, 732, 35\\

24a & Dessauges-Zavadsky, M., Calura, F., Prochaska, J.X., et al. 2004, A$\&$A, 416, 79 
& 31c & Balashev, S.A., Petitjean, P., Ivanchik, A.V., et al. 2011, MNRAS, 418, 357 \\

24c & Khare, P., Kulkarni, V.P., Lauroesch, J.T., et al. 2004, ApJ, 616, 86 
& 31P & Vladilo, G., Abate, C., Yin, J., et al. 2011, A$\&$A, 530, A33 \\ 

25a & Akerman, C.L., Ellison, et al. 2005, A$\&$A, 440, 499 
& 33a & Krogager, J.K., Fynbo, J.P.U., Ledoux, C., et al. 2013, MNRAS, 433, 3091 \\

25d & Srianand, R., Petitjean, P., Ledoux, C., et al. 2005, MNRAS, 362, 549 
& 34a & Vasquez, D.A., Rahamani, H., Noterdaeme, P., et al. 2014, A$\&$A, 562, A88\\

26a & Dessauges-Zavadsky, M., Prochaska, J.X., D'Odorico, S., et al. 2006, A$\&$A, 445, 93 
& 34B & Cooke, R., Pettini, M., et al. 2015, ApJ, 800, 12\\

26c & Herbert-Fort, S., Prochaska, J.X., Dessauges-Zavadsky, M., et al. 2006, PASP, 118, 1077 
& 34Z & Zafar, T., Centurión, M., Péroux, C., et al. 2014, MNRAS, 444, 744\\

26e & Ledoux, C., Petitjean, P., Fynbo, J.P.U., et al. 2006, A$\&$A, 457, 71 
& 35A & Quiret, S., Peroux, C., Zafar, T., et al. 2016, MNRAS\\

35B & Berg, T., Neeleman, M., Prochaska, X., et al. 2015, MNRAS, 452, 4326-4346 \\
\hline
\hline
\end{tabular}
\caption{References relative to the codes of Table~\ref{tab:DLA_sample}}  
\label{tab:references}
\end{center}
\end{table}

\bsp

\label{lastpage}

\end{document}